\documentclass[10pt,journal,compsoc]{IEEEtran}
\IEEEoverridecommandlockouts

\usepackage{caption2}
\usepackage{courier}
\usepackage{mathtools}
\usepackage{graphicx}
\usepackage{float}
\usepackage{subfigure}
\usepackage[ruled,linesnumbered]{algorithm2e}  
\usepackage{amsmath}  
\usepackage{booktabs}
\usepackage{multirow}
\usepackage{color}
\usepackage{amsfonts}
\usepackage{subfigure}
\usepackage{enumerate}
\usepackage{enumitem}
\setlist{leftmargin=3.5mm}
\usepackage{bm}
\newtheorem{definition}{Definition}
\newcommand{\tabitem}{~~\llap{\textbullet}~~}
\newcommand{\tabincell}[2]{\begin{tabular}{@{}#1@{}}#2\end{tabular}}  

\begin{document}

\title{\fontsize{22}{22}\selectfont A Survey on Heterogeneous Graph Embedding:\\Methods, Techniques, Applications and Sources}

\author{
		Xiao Wang,
		Deyu Bo,
		\IEEEauthorblockN{Chuan Shi\thanks{\IEEEauthorrefmark{2} Corresponding author}\IEEEauthorrefmark{2}},~\IEEEmembership{Member,~IEEE,}
		Shaohua Fan,
        Yanfang Ye,~\IEEEmembership{Member,~IEEE,}\\
        and Philip S. Yu,~\IEEEmembership{Fellow,~IEEE}
\IEEEcompsocitemizethanks{
\IEEEcompsocthanksitem X. Wang, D. Bo, C. Shi and S. Fan are with the Beijing Key Lab of Intelligent Telecommunications Software and Multimedia, Beijing University of Posts and Telecommunications, Beijing 100876, China. \protect\\
E-mail: \{xiaowang, bodeyu, shichuan, fanshaohua\}@bupt.edu.cn.
\IEEEcompsocthanksitem Y. Ye is with the Department
of Computer and Data Sciences, Case Western Reserve University, Cleveland,
OH, 44106.\protect\\
E-mail: yanfang.ye@case.edu.
\IEEEcompsocthanksitem P. S. Yu is with Computer Science Department, University of Illinois at
Chicago, Chicago, IL 60607, and also with the Institute for Data Science,
Tsinghua University, Beijing 100084, China. \protect
E-mail: psyu@uic.edu.}
}


\IEEEtitleabstractindextext{%
\begin{abstract}
	
Heterogeneous graphs (HGs) also known as heterogeneous information networks have become ubiquitous in real-world scenarios; therefore, HG embedding, which aims to learn representations in a lower-dimension space while preserving the heterogeneous structures and semantics for downstream tasks (e.g., node/graph classification, node clustering, link prediction), has drawn considerable attentions in recent years. In this survey, we perform a comprehensive review of the recent development on HG embedding methods and techniques. We first introduce the basic concepts of HG and discuss the unique challenges brought by the heterogeneity for HG embedding in comparison with homogeneous graph representation learning; and then we systemically survey and categorize the state-of-the-art HG embedding methods based on the information they used in the learning process to address the challenges posed by the HG heterogeneity. In particular, for each representative HG embedding method, we provide detailed introduction and further analyze its pros and cons; meanwhile, we also explore the transformativeness and applicability of different types of HG embedding methods in the real-world industrial environments for the first time. In addition, we further present several widely deployed systems that have demonstrated the success of HG embedding techniques in resolving real-world application problems with broader impacts. To facilitate future research and applications in this area, we also summarize the open-source code, existing graph learning platforms and benchmark datasets. Finally, we explore the additional issues and challenges of HG embedding and forecast the future research directions in this field.

\end{abstract}

}

\maketitle

\IEEEdisplaynontitleabstractindextext
\IEEEpeerreviewmaketitle

\IEEEraisesectionheading{\section{Introduction}\label{sec:introduction}}

\IEEEPARstart{H}{eterogeneous} graphs (HGs) \cite{sun2013mining}, which are capable of composing different types of entities (i.e., nodes) and relations, also known as heterogeneous information network, have become ubiquitous in real world scenarios, ranging from bibliographic networks, social networks to recommendation systems. For example, as shown in Fig. 1(a), a bibliographic network (i.e., academic network) can be represented by a HG, which consists of four types of nodes (author, paper, venue, and term) and three types of edges (author-write-paper, paper-contain-term and conference-publish-paper); and these basic relations can be further derived for more complex semantics over the HG (e.g., author-write-paper-contain-item). It has been well recognized that HG is a powerful model that is able to embrace rich semantics and structural information in real world data. Therefore, researches on HG data have been experiencing tremendous growth in data mining and machine learning, many of which have been successfully applied in real world systems such as recommendation \cite{shi2018heterogeneous, hu2018leveraging}, text analysis \cite{HGAT, GNUD}, and cybersecurity \cite{HACUD, hou2017hindroid}. 

Due to the ubiquity of HG data, how to learn embeddings of HG is a key research problem in various graph analysis applications, e.g., node/graph classification \cite{dong2017metapath2vec, fu2017hin2vec}, and node clustering \cite{li2019spectral}. Traditionally, to learn HG embeddings, matrix (e.g., adjacency matrix) factorization methods \cite{newman2006modularity, weisfeiler1968reduction} have been proposed to generate latent-dimension features in a HG. However, the computational cost of decomposing a large-scale matrix is usually very expensive, and also suffers from its statistical performance drawback \cite{shi2016survey, cui2018survey}. To address this challenge, heterogeneous graph embedding (i.e., heterogeneous graph representation learning), aiming to learn a function that maps input space into a lower-dimension space while preserving the heterogeneous structure and semantics, has drawn considerable attentions in recent years. Although there have been ample studies of embedding technology on homogeneous graphs \cite{cui2018survey} which consist of only one type of nodes and edges, these techniques cannot be directly applicable to HGs due to the heterogeneity of HG data. More specifically, 
i) the structure in HG is usually semantic dependent, e.g., meta-path structure \cite{dong2017metapath2vec}, implying that the local structure of one node in HG can be very different when considering different types of relations; 
ii) different types of nodes and edges have different attributes, which are usually located in different feature spaces, and thus when designing heterogeneous graph embedding methods, especially heterogeneous graph neural networks (HGNNs), we need to overcome the heterogeneity of attributes to fuse information \cite{wang2019heterogeneous, zhang2019heterogeneous};
iii) another one is that HG is usually application dependent: for example, the basic structure of HG usually can be captured by meta-path, however meta-path selection is still challenging in reality, which may need sufficient domain knowledge. 
To tackle the above issues, various heterogeneous graph embedding methods have been proposed \cite{chen2018pme, hu2019adversarial, dong2017metapath2vec, fu2017hin2vec, wang2019heterogeneous, shi2018heterogeneous}, many of which \cite{fan2019meta, zhao2019intentgc, HACUD, GEM, fan2018gotcha, ye2019out} have demonstrated the success of heterogeneous graph embedding techniques deployed in real world applications including recommendation systems \cite{shi2018heterogeneous, hu2018leveraging}, malware detection systems \cite{hou2017hindroid,fan2018gotcha,ye2019out,hou2019alphacyber}, and healthcare systems \cite{medical2020cao, anahita2018heteromed}. 

\begin{figure*}[!t]
	\centering 
	\includegraphics[width=\textwidth]{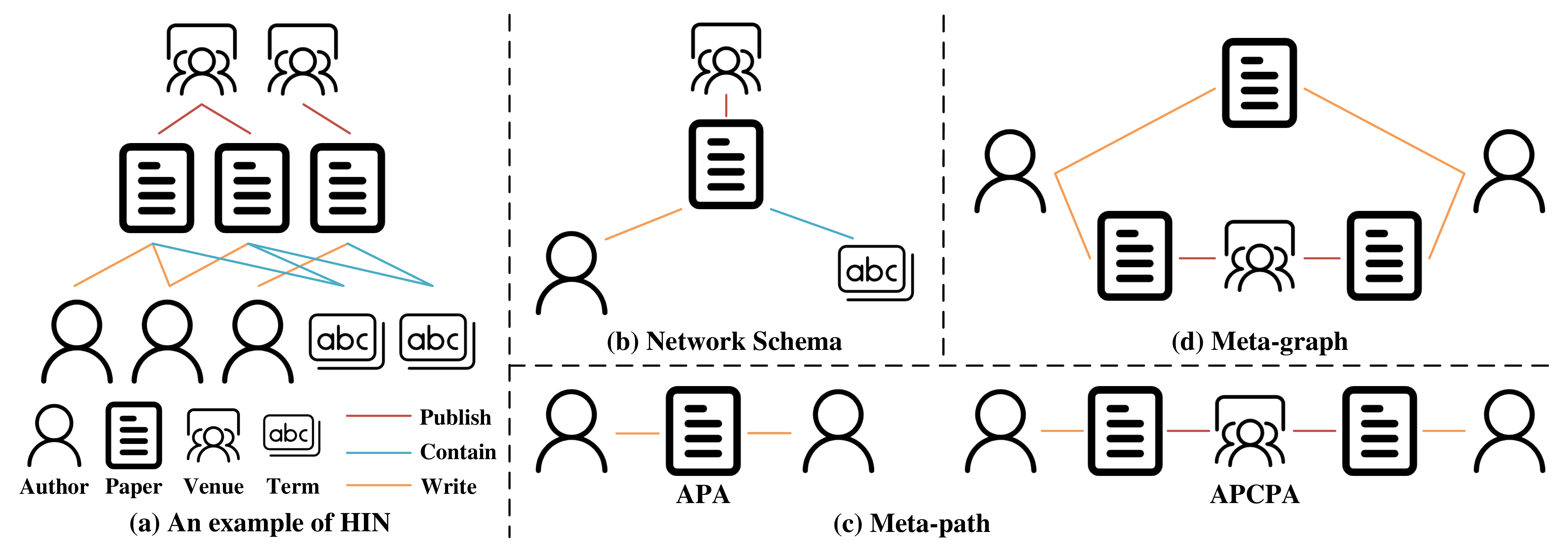}
	\caption{An illustrative example of a heterogeneous graph. (a) An academic network including four types of node (i.e., Author, Paper, Venue, Term) and three types of link (i.e., Publish, Contain, Write). (b) Network schema of the academic network. (c) Two meta-paths used in the academic network (i.e., Author-Paper-Author and Paper-Term-Paper). (d) A meta-graph used in the academic network.}
	\label{HIN_fig}
\end{figure*}

Although ample studies of heterogeneous graph embedding have been conducted with various applications in different fields, there have not been systematic and comprehensive survey on heterogeneous graph embedding methods with in-depth analysis of their pros and cons and detailed discussion of their transformativeness and applicability. To bridge this gap, in this paper, we will thoroughly survey the existing works on heterogeneous graph embedding, including representative methods and techniques, deployed systems in real world applications, and publicly available benchmark datasets as well as open-source code/tools. In particular, (1) we will explore recent progress of heterogeneous graph embedding, by introducing its representative methods and techniques with analysis of their pros and cons; then (2) we will introduce and discuss the transformativeness of existing heterogeneous graph embedding methods that have been successfully deployed in real-world applications; afterwards (3) we will summarize publicly available benchmark datasets and open-source code/tools to facilitate researchers and practitioners for future heterogeneous graph embedding works; and finally (4) we will discuss the additional issues and challenges of heterogeneous graph embedding technique and forecast the future research directions in this area. Note that different from the existing surveys which mainly focus on homogeneous graph embedding \cite{cui2018survey, zhang2018network, goyal2018graph, cai2018comprehensive, wu2020comprehensive, zhang2020deep}, we aim at exploring the works on heterogeneous graph embedding. Although there have been few survey works on heterogeneous graph embedding \cite{dong2020survey, yang2020survey}, we make our unique contributions in this work as summarized below. 

\begin{itemize}
	\item We first discuss the unique challenges brought by the heterogeneity of HG compared with homogeneous graphs; and then we provide a comprehensive survey of existing heterogeneous graph embedding methods, which are categorized based on the information they used in the learning process to address particular type of challenges posed by the HG heterogeneity.
    \item For each representative heterogeneous graph embedding method and technique, we provide detailed introduction and further analyze its pros and cons. In addition, we explore the transformativeness and applicability of different types of HG embedding methods in the real-world industrial environments for the first time.
	\item We summarize the open-source code and benchmark datasets, and give a detailed description to the existing graph learning platforms, to facilitate future research and applications in this area.
	\item We also explore the additional issues and challenges of heterogeneous graph embedding and forecast the future research directions in this field.
\end{itemize}

The remainder of this survey paper is organized as follows. In Section 2, we first introduce the HG concepts and discuss the unique challenges of heterogeneous graph embedding due to the heterogeneity. In Section 3, we categorize and introduce heterogeneous graph embedding methods in details according to the information (e.g., structures, attributes, and application dependent domain knowledge) used in the learning process, based on which we analyze their pros and cons and then discuss their applicability. In Section 4, we further summarize the commonly used techniques in the state-of-the-art heterogeneous graph embedding methods. In Section 5, we further explore the transformativeness of existing heterogeneous graph embedding methods that have been successfully deployed in real-world application systems. Section 5 summarizes the benchmark datasets and open-source code/tools for heterogeneous graph embedding. Section 7 discusses additional issues/challenges of heterogeneous graph embedding and forecasts the future research directions in this field. Finally, Section 8 concludes the paper.

\section{Preliminary}

In this section, we will first formally introduce the basic concepts in HG and illustrate the symbols used through this paper; and then we will elaborate the unique challenges brought by the heterogeneity of HG compared with homogeneous graphs.

\subsection{Basic Concepts}

HG is a graph consisting of different types of entities (i.e., nodes) and/or different types of relations (i.e., edges), which can be defined as follows.

\begin{definition}
\noindent
\textbf{Heterogeneous graph (or heterogeneous information network)} \cite{sun2013mining}. A HG is defined as a graph $\mathcal{G}=\{\mathcal{V}, \mathcal{E}\}$, in which $\mathcal{V}$ and $\mathcal{E}$ represent the node set and the link set, respectively. Each node $v \in \mathcal{V}$ and each link $e \in \mathcal{E}$ are associated with their mapping function $\phi(v): \mathcal{V} \to \mathcal{A}$ and $\varphi(e): \mathcal{E} \to \mathcal{R}$. $\mathcal{A}$ and $\mathcal{R}$ denote the node types and link types, respectively, where $\mathcal{A} + \mathcal{R} > 2$. The \textbf{network schema} for $\mathcal{G}$ is defined as $\mathcal{S}=(\mathcal{A}, \mathcal{R})$, which can be seen as a meta template of a heterogeneous graph $\mathcal{G}=\{\mathcal{V}, \mathcal{E}\}$ with the node type mapping function $\phi(v): \mathcal{V} \to \mathcal{A}$ and the link type mapping function $\varphi(e): \mathcal{E} \to \mathcal{R}$. The network schema is a graph defined over node types $\mathcal{A}$, with links as relations from $\mathcal{R}$.
\label{HIN}
\end{definition}

HG not only provides the graph structure of the data associations, but also provides a higher-level semantics of the data. An example of HG is illustrated in Fig. \ref{HIN_fig}(a), which consists of four node types (author, paper, venue, and term) and three link types (author-write-paper, paper-contain-term, and conference-publish-paper); while Fig. \ref{HIN_fig}(b) illustrates the network schema. Based on a constructed HG, to formulate the semantics of higher-order relationships among entities, meta-path \cite{sun2011pathsim} is further proposed whose definition is given below. 

\begin{definition}
\textbf{Meta-path} \cite{sun2011pathsim}. A meta-path $m$ is based on a network schema $\mathcal{S}$, which is denoted as $m=A_{1}\stackrel{R_{1}}{\longrightarrow}A_{2}\stackrel{R_{2}}{\longrightarrow} \cdots \stackrel{R_{l}}{\longrightarrow}A_{l+1}$ (simplified to $A_{1}A_{2} \cdots A_{l+1}$) with node types $A_{1}, A_{2}, \cdots ,A_{l+1} \in \mathcal{A}$ and link types $R_{1}, R_{2}, \cdots R_{l} \in \mathcal{R}$.
\label{meta-path}
\end{definition}

Note that different meta-paths describe semantic relationships in different views. For example, the meta-path of ``APA'' indicates the co-author relationship and ``APCPA'' represents the co-conference relation. Both of them can be used to formulate the relatedness over authors. Although meta-path can be used to depict the relatedness over entities, it fails to capture a more complex relationship, such as motifs \cite{milo2002network}. To address this challenge, meta-graph \cite{huang2016meta} is proposed to use a directed acyclic graph of entity and relation types to capture more complex relationship between two HG entities, defined as follows.

\begin{definition}
\textbf{Meta-graph} \cite{huang2016meta}. A meta-graph $\mathcal{T}$ can be seen as a directed acyclic graph (DAG) composed of multiple meta-paths with common nodes.
Formally, meta-graph is defined as $\mathcal{T}=(V_{\mathcal{T}}, E_{\mathcal{T}})$, where $V_{\mathcal{T}}$ is a set of nodes and $E_{\mathcal{T}}$ is a set of links. For any node $v \in V_{\mathcal{T}}, \phi(v) \in \mathcal{A}$; for any link $e \in E_{\mathcal{T}}, \varphi(e) \in \mathcal{R}$.
\label{metagraph}
\end{definition}

An example meta-graph is shown in Fig. \ref{HIN}(d), which can be regarded as the combination of meta-path ``APA'' and ``APCPA'', reflecting a high-order similarity of two nodes. Note that a meta-graph can be symmetric or asymmetric \cite{mg2vec}. To learn embeddings of HG data, we formalize the problem of heterogeneous graph embedding as follow.

\begin{definition}
\textbf{Heterogeneous graph embedding} \cite{shi2016survey}. Heterogeneous graph embedding aims to learn a function $\Phi: \mathcal{V} \to \mathbb{R}^{d}$ that embeds the nodes $v \in \mathcal{V}$ in HG into a low-dimensional Euclidean space with $d \ll |\mathcal{V}|$.
\label{HIN embedding}
\end{definition}
Table 1 summarizes symbols used through this paper. 

\begin{table}[htbp]
  \centering
  \caption{Notations and Explanations}
    \begin{tabular}{cc}
    \toprule
    Notations & Explainations \\
    \midrule
	$d$						& dimension of node embeddings \\
	$N$						& Number of nodes \\
	$m$						& Meta-path\\
	$\mathbf{h}_{i}$		& Attributes or embeddings of node $i$ \\
    $\mathbf{M}_{r}$		& Relation-specific matrix of relation $r$ \\
	$w_{ij}$				& Weight of link between node $i$ and node $j$ \\
 	$S_{r}$					& Heterogeneous similarity function with relation $r$ \\
    $C_{t}(i)$				& Context nodes of node $i$ with type $t$ \\
 	$\mathcal{N}_{i}$	& Neighbors of node $i$ \\
 	$\sigma$				& Sigmoid function \\
 	$\odot$					& Hadamard product \\
 	$\oplus$				& Concatenation operator\\
    \bottomrule
    \end{tabular}
  \label{notations}
\end{table}

\subsection{Challenges of HG Embedding due to Heterogeneity}
\label{challenges}

Different from homogeneous graph embedding \cite{cui2018survey}, where the basic problem is preserving structure and property in node embedding \cite{cui2018survey}. Due to the heterogeneity, heterogeneous graph embedding imposes more challenges, which are illustrated below.

\begin{figure*}[t]
	\centering
	\includegraphics[width=\textwidth]{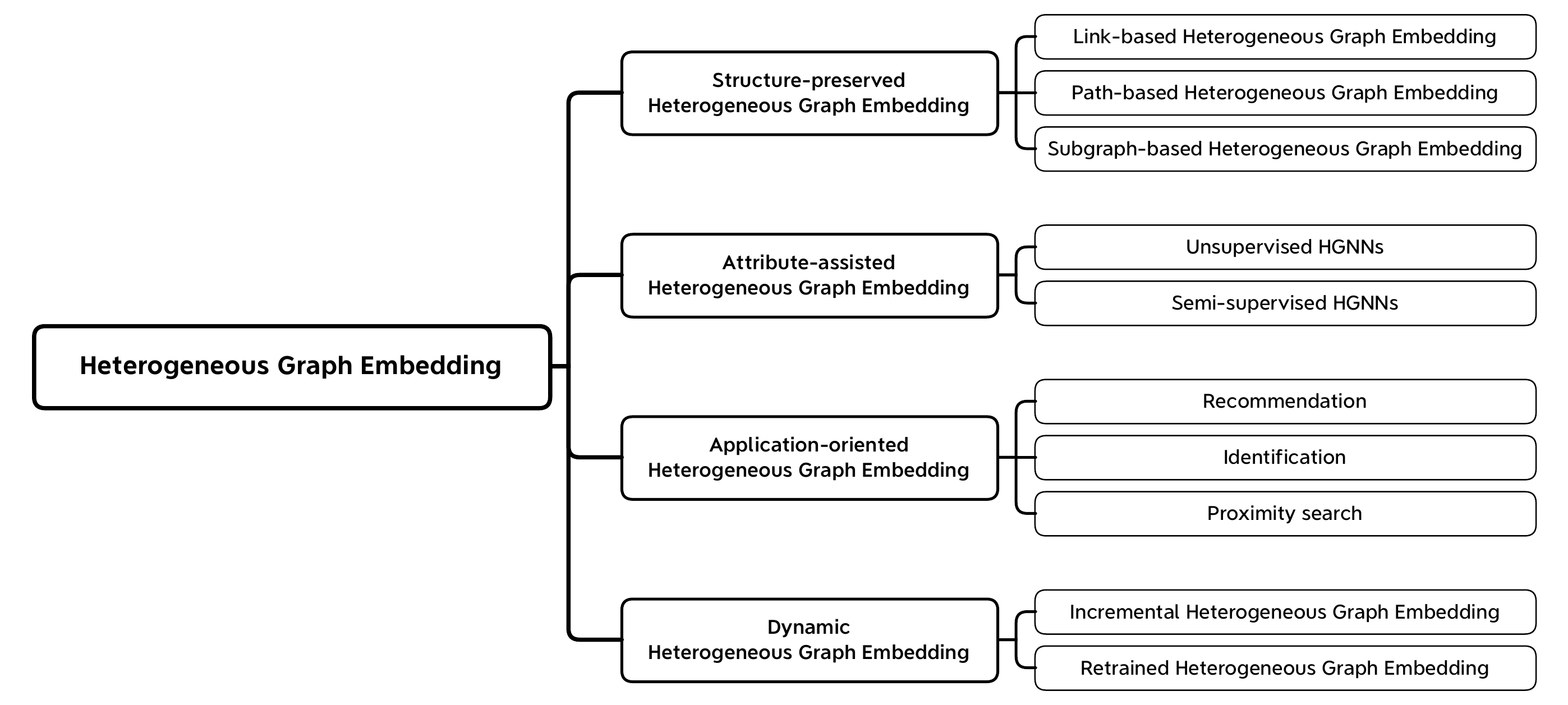}
	\caption{An overview of heterogeneous graph embedding from the perspective of used information.}
	\label{overview}
\end{figure*}

\begin{itemize}
	\item \textbf{Complex structure} (the complex HG structure caused by multiple types of nodes and edges). In a homogeneous graph, the fundamental structure can be considered as the so-called first-order, second-order, and even higher-order structure \cite{perozzi2014deepwalk, tang2015line, wang2017community}. All these structures are well defined and have good intuition. However, the structure in HG will dramatically change depending on the selected relations. Let's still take the academic network in Fig. 1(a) as an example, the neighbors of one paper will be authors with the ``write" relation, while with ``contain" relation, the neighbors become terms. Complicating things further, the combination of these relations, which can be considered as a higher-order structure in HG, will result in different and more complicated structures. Therefore, how to efficiently and effectively preserve these complex structures is of great challenge in heterogeneous graph embedding, while current efforts have been made towards the meta-path structure \cite{dong2017metapath2vec} and meta-graph structure \cite{zhang2018metagraph2vec}, etc.
	
	\item \textbf{Heterogeneous attributes} (the fusion problem caused by the heterogeneity of attributes). Since the nodes and edges in a homogeneous graph have the same type, each dimension of the node or edge attributes has the same meaning. In this situation, node can directly fuse the attributes of its neighbors. However, in heterogeneous graph, the attributes of different types of nodes and edges may have different meanings \cite{zhang2019heterogeneous, wang2019heterogeneous}. For example, the attributes of author can be the research fields, while paper may use keywords as attributes. Therefore, how to overcome the heterogeneity of attributes and effectively fuse the attributes of neighbors poses another challenge in heterogeneous graph embedding.
	
	\item \textbf{Application  dependent}. HG is closely related to the real world applications, while many practical problems remain unsolved. For example, constructing an appropriate HG may require sufficient domain knowledge in a real world application. Also, meta-path and/or meta-graph are widely used to capture the structure of HG. However, unlike homogeneous graph, where the structure (e.g., the first-order and second-order structure) is well defined, meta-path selection may also need prior knowledge. Furthermore, to better facilitate the real world applications, we usually need to elaborately encode the side information (e.g., node attributes) \cite{wang2019heterogeneous, zhang2019heterogeneous, DyHNE, DyHAN} or more advanced domain knowledge \cite{shi2018heterogeneous, chen2017task, liu2018interactive} to the heterogeneous graph embedding process. 
	
\end{itemize}

\section{Method Taxonomy}
\label{taxonomy}

Various types of nodes and links in HG bring different graph structures and rich attributes (i.e., heterogeneity). As discussed in Section \ref{challenges}, in order to make the node embeddings capture the heterogeneous structures and rich attributes, we need to consider the information of different aspects in the embedding, including graph structures, attributes and specific application labels, etc.
Based on the aforementioned challenges, in this section, we categorize the existing methods into four categories based on the information they used in heterogeneous graph embedding:
(1) \textit{Structure-preserved heterogeneous graph embedding}. The methods belonging to this category primarily focus on capturing and preserving the heterogeneous structures and semantics, e.g., the meta-path and meta-graph.
(2) \textit{Attribute-assisted heterogeneous graph embedding}. The methods incorporate more information beyond structure, e.g., node and edge attributes, into embedding technology, so as to utilize the neighborhood information more effectively.
(3) \textit{Application-oriented heterogeneous graph embedding}. We further explore the applicability of the heterogeneous graph embedding methods (i.e., the ones aim to learn application-oriented node embeddings over HG). 
(4) \textit{Dynamic heterogeneous graph embedding}. Different from existing survey works that mainly focus on the embedding methods for static heterogeneous graphs. In this work, we further explore and summarize dynamic heterogeneous graph embedding methods, which aim to capture the evolution of heterogeneous graphs and preserve the temporal information in the node embeddings.
An overview of different types of heterogeneous graph embedding methods explored in this survey paper is shown in Fig. \ref{overview}.

\subsection{Structure-preserved HG Embedding}
\label{structure}
One basic requirement of graph embedding is to preserve the graph structures properly \cite{cui2018survey}. Accordingly, the homogeneous graph embedding pays more attention on higher-order graph structures, for example, second-order structures \cite{tang2015line, wang2016structural}, high-order structures \cite{graphrep, AROPE} and community structures \cite{wang2017community}.
However, one typical characteristic of HG is that it contains multiple relations among nodes, which inevitably needs to consider the heterogeneity of graph. Therefore, an important direction of heterogeneous graph embedding is to learn semantic information from the graph structures. In this section, we review the typical heterogeneous graph embedding methods based on the basic HG structures, including link (i.e., edge), meta-path, and subgraph. Link is the observed relation between two nodes, meta-path is composed of different types of links and subgraph represents the tiny sub-structure of graph. The three structures are the most fundamental ingredients of HG, which are able to capture the semantic information from different perspectives. In the followings, we will review the typical structure-preserved heterogeneous graph embedding methods based on these three types of structures and then discuss their pros and cons.

\subsubsection{Link-based HG Embedding}

One of the most basic information that heterogeneous graph embedding needs to preserve is link. Different from homogeneous graph, link in HG has different types and contains different semantics. To distinguish various types of links, one classical idea is to deal with them in different metric spaces, rather than a unified metric space. A representative work is PME \cite{chen2018pme}, which treats each link type as a relation and uses a relation-specific matrix to transform the nodes into different metric spaces. In this way, nodes connected by different types of links can be close to each other in different metric spaces, thus capturing the heterogeneity of the graph. The distance function is defined as follows:
\begin{equation}
	S_{r}(v_{i},v_{j})=w_{ij}\left\| \mathbf{M}_{r} \mathbf{h}_{i} - \mathbf{M}_{r} \mathbf{h}_{j} \right\|_{2},
\label{eq_PME}
\end{equation}
where $\mathbf{h}_{i}$ and $\mathbf{h}_{j} \in \mathbb{R}^{d \times 1}$ denote the node embeddings of node $i$ and node $j$, respectively; $\mathbf{M}_{r} \in \mathbb{R}^{d \times d}$ is the projection matrix of relation $r$; and $w_{ij}$ represents the weight of link between node $i$ and node $j$. Note that Eq. \ref{eq_PME} can be seen as a metric learning function:
\begin{equation}
	\left\| \mathbf{M}_{r} ( \mathbf{h}_{i} - \mathbf{h}_{j} ) \right\|_{2} = \sqrt{ ( \mathbf{h}_{i} - \mathbf{h}_{j} )^\top \mathbf{M}_{r}^\top \mathbf{M}_{r} ( \mathbf{h}_{i} - \mathbf{h}_{j} )},
\label{eq_metric}
\end{equation}
where $\mathbf{M}_{r}^\top \mathbf{M}_{r} \in \mathbb{R}^{d \times d}$ is the metric matrix of Mahalanobis distance \cite{metric}. PME considers the relations between nodes when minimizing the distance of them, thus capturing the heterogeneity of graph. The loss function is the margin-based triple loss function, which requires a distance between the positive and negative samples:
\begin{equation}
	L=\sum_{r \in \mathcal{R}}\sum_{(v_{i},v_{j}) \in E_{r}}\sum_{(v_{i},v_{k}) \notin E_{r}}[\xi + S_{r}(v_{i},v_{j})^2 - S_{r}(v_{i},v_{k})^2]_{+}
	\label{loss_PME}
\end{equation}
where $\xi$ denotes the margin, $E_{r}$ represents the positive links of relation $r$, and $[z]_{+}=\max(z,0)$. Through Eq. \ref{loss_PME}, PME makes the node pairs connected by relation $r$ closer to each other than the node pairs without relation $r$.

By exploiting the relation-specific matrix to capture the link heterogeneity, different from PME, other methods have been proposed aiming to maximize the similarity of two nodes connected by specific relations. For example, EOE \cite{xu2017embedding} and HeGAN \cite{hu2019adversarial} use the relation-specific matrix $\mathbf{M}_{r}$ to calculate the similarity between two nodes, which can be formulated as:
\begin{equation}
\label{EOE}
	S_{r}(v_{i}, v_{j})=\frac{1}{1+\exp\left\{ -\mathbf{h}^{\top}_{i} \mathbf{M}_{r} \mathbf{h}_{j} \right\}}.
\end{equation}
More specifically, EOE is proposed to learn embeddings for coupled heterogeneous graphs, which consist of two different but related sub-graphs. It divides the links in HG into intra-graph links and inter-graph links. Intra-graph link connects two nodes with the same type, and inter-graph link connects two nodes with different types. To capture the heterogeneity in inter-graph link, EOE utilizes Eq. \ref{EOE} as the similarity function of two nodes. Different from EOE, HeGAN uses generative adversarial networks (GAN) \cite{GAN} to learn node embeddings for heterogeneous graph. It uses Eq. \ref{EOE} as a discriminator to determine whether the node embeddings are produced by the generator. Through the game between discriminator and generator, HeGAN can learn more robust node embeddings.

The previously discussed methods mainly preserve the link structure based on either the distance or similarity function on node embeddings, while AspEM \cite{shi2018aspem} and HEER \cite{shi2018easing} aim to maximize the probability of existing links. The heterogeneous similarity function is defined as:
\begin{equation}
	S_{r}=\frac{\exp(\bm{\mu}^{\top}_{r}\mathbf{g}_{ij})}{\sum_{\tilde{i} \in E^{r}_{\tilde{i}j}} \exp(\bm{\mu}^{\top}_{r}\mathbf{g}_{\tilde{i}j}) + \sum_{\tilde{j} \in E^{r}_{i\tilde{j}}} \exp(\bm{\mu}^{\top}_{r}\mathbf{g}_{i\tilde{j}})},
\end{equation}
where $\bm{\mu}_{r} \in \mathbb{R}^{d \times 1}$ is the embedding of relation $r$; $\mathbf{g}_{ij} \in \mathbb{R}^{d \times 1}$ is the embedding of link between node $i$ and node $j$; $\mathbf{g}_{ij} = \mathbf{h}_{i} \odot \mathbf{h}_{j}$ and $\odot$ denotes the Hadamard product; and $E^{r}_{\tilde{i}j}$ is the set of negative links, which indicates that there is no link between node $\tilde{i}$ and node $j$. 
It can be seen that $\bm{\mu}^{\top}_{r}\mathbf{g}_{ij}$ measures the closeness between link and its corresponding type. Maximizing $S_{r}$ enlarges the closeness between the existing links and their corresponding types, thus capturing the heterogeneity of the graph.

In addition to the above methods, there are some methods that draw on techniques from other fields. Similar to the idea of TransE \cite{bordes2013translating}, MELL \cite{matsuno2018mell} uses the equation 'head + relation = tail' to learn the node embeddings for heterogeneous graph. PTE \cite{tang2015pte} decomposes the heterogeneous graph into multiple bipartite graphs and employs LINE \cite{tang2015line}, which preserves the first- and second-order structures of graph, to learn node embeddings for each bipartite graph. MNE \cite{zhang2018scalable} assigns multiple embeddings for each node and uses a skip-gram technique \cite{mikolov2013distributed} to represent information of multi-type relations into a unified space.

In summary, we can roughly divide the link-based heterogeneous graph embedding methods into two categories: one is to explicitly preserve the proximity of links \cite{shi2018aspem, shi2018easing}; the other is to preserve the proximity of nodes, which utilizes the information of links implicitly \cite{chen2018pme, xu2017embedding, hu2019adversarial}. These two types of methods both make use of the first-order information of HG.

\subsubsection{Path-based HG Embedding}

Link-based methods can only capture the local structures of HG, i.e., the first-order relation. In fact, the higher-order relation, describing more complex semantic information, is also critical for heterogeneous graph embedding.
For example, in Fig. \ref{HIN}(a), the first-order relation can only reflect the similarity of author-paper, paper-term and paper-venue. While the similarity of author-author, paper-paper and author-conference cannot be well captured. Therefore, the high-order relation is introduced to measure more complex similarity. Because the number of high-order relations is very large, in order to reduce complexity, we usually choose the higher-order relations with rich semantics, called meta-path. In this section, we will introduce some representative meta-path-based heterogeneous graph embedding methods, which can be divided into two categories: random walk-based methods \cite{dong2017metapath2vec, he2019hetespaceywalk, hussein2018meta, lee2019bhin2vec, wang2019hyperbolic} and hybrid relation-based methods \cite{fu2017hin2vec, lu2019relation}.

\begin{figure}[ht]
	\centering
	\includegraphics[scale=0.5]{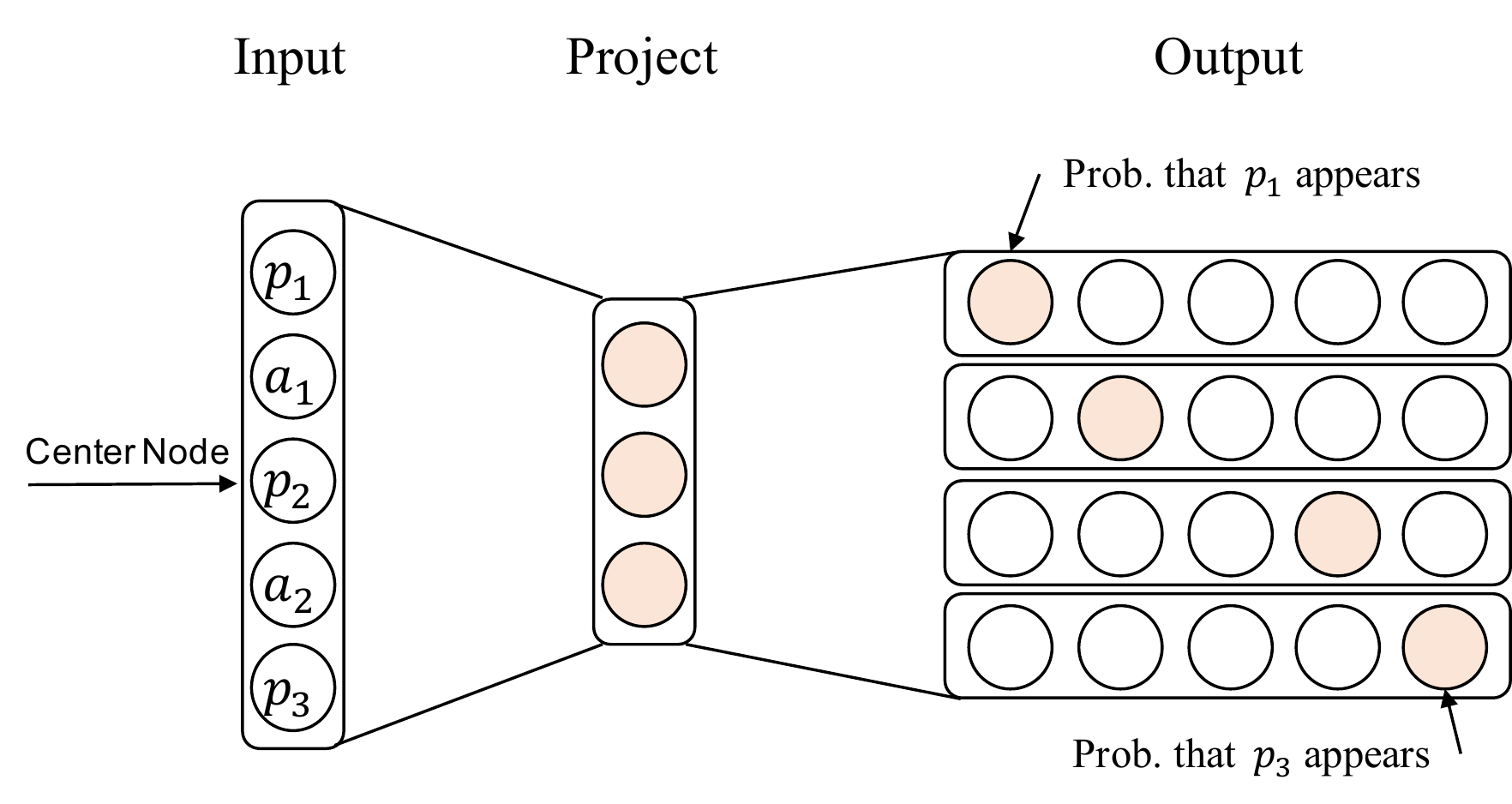}
	\caption{The architecture of metapath2vec. Node sequence is generated under the meta-path PAP. It projects the embedding of the center node, e.g., $p_{2}$ into latent space and maximizes the probability of its meta-path-based context nodes, e.g., $p_{1}$, $p_{3}$, $a_{1}$ and $a_{2}$, appearing.}
	\label{mp2v}
\end{figure}

Random walk-based methods usually use meta-path to guide random walk on a HG, so that the generated node sequence contains rich semantic information. Through preserving the node sequence structure, node embedding can preserve both first-order and high-order proximity \cite{dong2017metapath2vec}. A representative work is  metapath2vec \cite{dong2017metapath2vec} (shown in Fig. \ref{mp2v}).

Metapath2vec \cite{dong2017metapath2vec} mainly uses meta-path guided random walk to generate heterogeneous node sequences with rich semantics; and then it designs a heterogeneous skip-gram technique to preserve the proximity between node $v$ and its context nodes, i.e., neighbors in the random walk sequences:
\begin{equation}
\mathop{\arg}\max_{\theta} \sum_{v \in \mathcal{V}} \sum_{t \in \mathcal{A}} \sum_{c_{t} \in C_{t}(v)} \log p(c_{t}|v;\theta),
\label{mpobj}
\end{equation}
where $C_{t}(v)$ represents the context nodes of node $v$ with type $t$. $p(c_{t}|v;\theta)$ denotes the heterogeneous similarity function on node $v$ and its context neighbors $c_{t}$:
\begin{equation}
p(c_{t}|v;\theta)=\frac{e^{\mathbf{h}_{c_{t}} \cdot \mathbf{h}_{v}}}{\sum_{\tilde{v} \in \mathcal{V}}e^{\mathbf{h}_{\tilde{v}} \cdot \mathbf{h}_{v}}},
\label{mpsim}
\end{equation}

From the diagram shown in Fig. \ref{mp2v}, Eq. \ref{mpsim} needs to calculate the similarity between center node and its neighbors. Then \cite{mikolov2013distributed} introduces a negative sampling strategy to reduce the computation. Hence, Eq. \ref{mpsim} can be approximated as:
\begin{equation}
	\log\sigma(\mathbf{h}_{c_{t}} \cdot \mathbf{h}_{v}) + \sum_{q=1}^{Q} \mathbb{E}_{\tilde{v}^{q} \sim P(\tilde{v})} \left[ \log\sigma \left(- \mathbf{h}_{\tilde{v}^{q}} \cdot \mathbf{h}_{v} \right) \right],
\end{equation}
where $\sigma(\cdot)$ is the sigmoid function, and $P(\tilde{v})$ is the distribution in which the negative node $\tilde{v}^{q}$ is sampled for $Q$ times. Through the strategy of negative sampling, the time complexity is greatly reduced. However, when choosing the negative samples, metapath2vec does not consider the types of nodes, i.e., different types of nodes are from the same distribution $P(\tilde{v})$. It further designs metapath2vec++, which samples the negative nodes of the same type as the central node, i.e., $\tilde{v}_{t}^{q} \sim P(\tilde{v}_{t})$. The formulation can be rewritten as:
\begin{equation}
	\log\sigma(\mathbf{h}_{c_{t}} \cdot \mathbf{h}_{v}) + \sum_{q=1}^{Q} \mathbb{E}_{\tilde{v}_{t}^{q} \sim P(\tilde{v}_{t})} \left[ \log\sigma \left(- \mathbf{h}_{\tilde{v}_{t}^{q}} \cdot \mathbf{h}_{v} \right) \right].
\end{equation}
After minimizing the objective function, metapath2vec and metapath2vec++ can capture both structural information and semantic information effectively and efficiently.

Based on metapath2vec, a series of variants have been proposed.
Spacey \cite{he2019hetespaceywalk} designs a heterogeneous spacey random walk to unify different meta-paths with a second-order hyper-matrix to control the transition probability among different node types. JUST \cite{hussein2018meta} proposes a random walk method with Jump and Stay strategies, which can flexibly choose to change or maintain the type of the next node in the random walk without meta-path. BHIN2vec \cite{lee2019bhin2vec} proposes an extended skip-gram technique to balance the various types of relations. It treats heterogeneous graph embedding as a multiple relation-based tasks, and balances the influence of different relations on node embeddings by adjusting the training ratio of different tasks. HHNE \cite{wang2019hyperbolic} conducts the meta-path guided random walk in hyperbolic spaces \cite{helgason1979differential}, where the similarity between nodes can be measured using hyperbolic distance. In this way, some properties of HG, e.g., hierarchical and power-law structure, can be naturally reflected in the learned node embeddings.

\begin{figure}[ht]
\centering
\includegraphics[width=\linewidth]{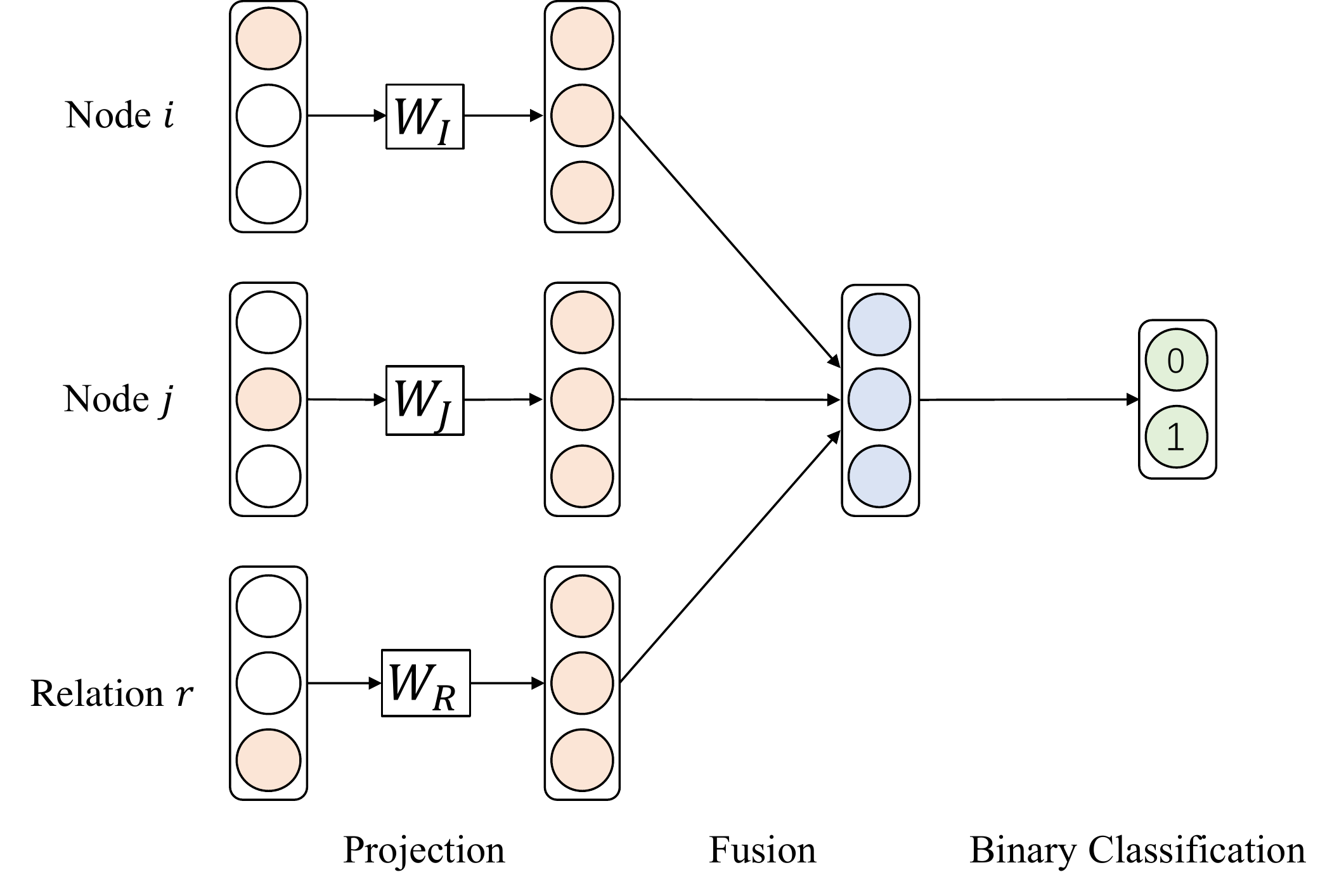}
\caption{The architecture of HIN2vec. Through the parameter matrices $\mathbf{W}_{I}, \mathbf{W}_{J}$ and $\mathbf{W}_{R}$, the one-hot vectors of node $i$, node $j$ and relation $r$ are projected into dense vectors. And these three vectors are fused by a neural network to predict whether node $i$ and $j$ are connected by relation $r$.}
\label{HIN2vec}
\end{figure}

Different from random walk-based methods that learn structural and semantic information from generated node sequences, some methods use the combination of first-order relation and high-order relation (i.e., meta-path) to capture the heterogeneity of HG. We call these work as hybrid relation-based methods. A typical work is HIN2vec \cite{fu2017hin2vec} (shown in Fig. \ref{HIN2vec}), which carries out multiple relation prediction tasks jointly to learn the embeddings of nodes and meta-paths.

The purpose of HIN2vec is to predict whether two nodes are connected by a meta-path, which can be seen as a multi-label classification task. As illustrated in Fig. \ref{HIN2vec}, given two nodes $i$ and $j$, HIN2vec uses the following function to compute their similarity under the hybrid relation $r$:
\begin{equation}
	S_{r}(v_i, v_j)=\sigma(\sum \mathbf{W}_{I}\vec{i} \odot \mathbf{W}_{J}\vec{j} \odot f_{01}(\mathbf{W}_{R}\vec{r})),
\end{equation}
where $\vec{i}, \vec{j}$ and $\vec{r} \in \mathbb{R}^{N \times 1}$ denote the one-hot vectors of nodes and relation, respectively; $\mathbf{W}_{I}, \mathbf{W}_{J}$ and $\mathbf{W}_{R} \in \mathbb{R}^{d \times N}$ are the mapping matrices; and $f_{01}(\cdot)$ is a regularization function, which limits the embedding values between 0 and 1. The loss function is a binary cross-entropy loss:
\begin{equation}
	E^{r}_{ij}\log S_{r}(v_i, v_j) + [1-E^{r}_{ij}]\log [1-S_{r}(v_i, v_j)],
\end{equation}
where $E^{r}_{ij}$ denotes the set of positive links. After minimizing the loss function, HIN2vec can learn the embedding of nodes and relations (meta-paths). Besides, in the relation set $R$, it contains not only the first-order structures (e.g., A-P relation) but also the high-order structures (e.g., A-P-A relation). Therefore, the node embeddings can capture different semantics.

RHINE \cite{lu2019relation} is another hybrid relation-based method, which designs different distance functions for different relations, thus enhancing the expressive power of node embeddings. It divides the relations into two categories: Affiliation Relations (ARs) and Interaction Relations (IRs). For ARs, a Euclidean distance function is introduced; while for IRs, RHINE proposes a translation-based distance function. Through the combination of these two distance functions, RHINE can learn relation structure-aware heterogeneous node embeddings.

In sum, we can find that random walk-based methods mainly exploit meta-path guided strategy for heterogeneous graph embedding; while hybrid relation-based methods regard a meta-path as high-order relation and learn meta-path based embeddings simultaneously. Compared with random walk-based methods, hybrid relation-based methods can simultaneously integrate multiple meta-paths into heterogeneous graph embedding flexibly.

\subsubsection{Subgraph-based HG Embedding}

Subgraph represents a more complex structure in the graph. Incorporating subgraphs into graph embedding can significantly improve the ability of capturing complex structural relationships. In this section, we introduce two widely used subgraphs in HG: one is metagraph, which reflects the high-order similarity between nodes \cite{zhang2018metagraph2vec, mg2vec}; the other is the hyperedge\footnotemark[1], which connects a series of closely related nodes and preserves the indecomposablity among them \cite{tu2018structural}.

\footnotetext[1]{In this paper, we treat the hyperedge as a special kind of subgraph.}

Zhang \emph{et al.} propose metagraph2vec \cite{zhang2018metagraph2vec}, which uses a metagraph-guided random walk to generate heterogeneous node sequence. 
Then the heterogeneous skip-gram technique \cite{dong2017metapath2vec} is employed to learn the node embeddings. Based on this strategy, metagraph2vec can capture the rich structural information and high-order similarity among nodes.
Different from metagraph2vec that only uses metagraphs in the pre-processing step (i.e., metagraph-guided random walk), mg2vec \cite{mg2vec} aims to learn the embeddings for metagraphs and nodes jointly, so that the metagraphs can join the learning process. It first enumerates the metagraphs and then preserves the proximity between nodes and metagraphs:
\begin{equation}
	P(\mathbf{M}_{i}|v)=\frac{\exp (\mathbf{M}_{i} \cdot \mathbf{h}_{v})}{\sum_{M_{j} \in \mathcal{M}} \exp (\mathbf{M}_{j} \cdot \mathbf{h}_{v})},
\end{equation}
where $\mathbf{M}_{i}$ is the embedding of metagraph $i$ and $\mathcal{M}$ denotes the set of metagraphs. Clearly, $P(M_{i}|v)$ represents the first-order relationship between the nodes and its subgraphs. Further, mg2vec preserves the proximity between node pair and its subgraph to capture the second-order information:
\begin{equation}
	P(\mathbf{M}_{i}|u, v)=\frac{\exp (\mathbf{M}_{i} \cdot f(\mathbf{h}_{u}, \mathbf{h}_{v}))}{\sum_{M_{j} \in \mathcal{M}} \exp (\mathbf{M}_{j} \cdot f(\mathbf{h}_{u}, \mathbf{h}_{v}))},
\end{equation}
where $f(\cdot)$ is a neural network to learn the embeddings of node pairs.
Through preserving the first-order and second-order proximity between nodes and metagraphs, mg2vec can capture the structural information and the similarity between nodes and metagraphs.

DHNE \cite{tu2018structural} is a typical hyperedge-based graph embedding method. Specifically, it designs a novel deep model to produce a non-linear tuple-wise similarity function while capturing the local and global structures of a given HG. Taking a hyperedge with three nodes $a, b$ and $c$ as an example. The first layer of DHNE is an autoencoder, which is used to learn latent embeddings and preserve the second-order structures of graph \cite{tang2015line}. The second layer is a fully connected layer with embedding concatenated:
\begin{equation}
	\mathbf{L}=\sigma(\mathbf{W}_{a}\mathbf{h}_{a} \oplus \mathbf{W}_{b}\mathbf{h}_{b} \oplus \mathbf{W}_{c}\mathbf{h}_{c}),
\end{equation}
where $\mathbf{L}$ denotes the embedding of the hyperedge; $\mathbf{h}_{a}, \mathbf{h}_{b}$ and $\mathbf{h}_{c} \in \mathbb{R}^{d \times 1}$ are the embeddings of node $a$, $b$ and $c$ learn by the autoencoder. $\mathbf{W}_{a}, \mathbf{W}_{b}$ and $\mathbf{W}_{c} \in \mathbb{R}^{d' \times d}$ are the transformation matrices for different node types. Finally, the third layer is used to calculate the indecomposability of the hyperedge:
\begin{equation}
	\mathcal{P}=\sigma(\mathbf{W}*\mathbf{L}+\mathbf{b}),
\end{equation}
where $\mathcal{P}$ denote the indecomposability of the hyperedge; $\mathbf{W} \in \mathbb{R}^{1 \times 3d'}$ and $\mathbf{b} \in \mathbb{R}^{1 \times 1}$ are the weight matrix and bias, respectively. A higher value of $\mathcal{P}$ means these nodes are from the existing hyperedges, otherwise it should be small.
HEBE \cite{gui2016large} is another hyperedge-based method, which aims to maximize the proximity between the node and the hyperedge it belongs to. After maximizing the proximity, HEBE can preserve the similarity of nodes within the same hyperedge, while reduce the similarity of nodes from different hyperedges. Besides, \cite{huang2019hyperpath} proposes hyper-path-based random walk to preserve both the structural information and indecomposability of the hyper-graphs.

Compared with the structures of link and meta-path, subgraph (with two representative forms of meta-graph and hyperedge) usually contains much higher order structural and semantic information. However, one obstacle of subgraph-based heterogeneous graph embedding methods is the high complexity of subgraph. How to balance the effectiveness and efficiency is required for a practical subgraph-based heterogeneous graph embedding methods, which is worthy of further exploration.

\subsubsection{Summary}

Generally, structure-preserved heterogeneous graph embedding methods mainly use shallow models, i.e., models without non-linear activation and multiple transformation. A major advantage of this type of methods is that they have good parallelizability and can improve training speed through negative sampling \cite{mikolov2013distributed}. However, as we can seen, there has been increasingly advanced structural and semantic information from link to path to subgraph, which may improve the performance in nature, but it also requires more calculations. Besides, there are two serious problems: one is that the shallow models need to assign each node a low-dimensional embedding, which requires larger memory spaces to store the parameters. Another is that shallow models can only work on transductive setting, i.e., they cannot learn the embedding of new node. These two shortcomings limit the application of this kind of methods in large-scale industrial scenarios.


\subsection{Attribute-assisted HG Embedding}
\label{side}

In addition to the graph structures, another important component of heterogeneous graph embedding is the rich attributes. Attribute-assisted heterogeneous graph embedding methods aim to encode the complex structures and multiple attributes to learn node embeddings.
Different from graph neural networks (GNNs) that can directly fuse the attributes of neighbors to update node embeddings, due to the different types of nodes and edges, HGNNs need to overcome the heterogeneity of attributes and design effective fusion methods to utilize the neighborhood information, thus bringing more challenges.
In this section, we divide HGNNs into unsupervised and semi-supervised settings, then discuss their pros and cons.

\subsubsection{Unsupervised HGNNs}

Unsupervised HGNNs aim to learn node embeddings with good generalization. To this end, they always utilize the interactions among different types of attributes to capture the potential commonalities.

HetGNN \cite{zhang2019heterogeneous} is the representative work of unsupervised HGNNs. It consists of three parts: content aggregation, neighbor aggregation and type aggregation.
Content aggregation is designed to learn fused embeddings from different node contents, such as images, text or attributes:
\begin{equation}
	f_{1}(v)=\frac{\sum_{i \in C_{v}}[ \overrightarrow{LSTM} \{ \mathcal{FC}(\mathbf{h}_{i}) \} \oplus \overleftarrow{LSTM} \{ \mathcal{FC}(\mathbf{h}_{i}) \} ]}{|C_{v}|},
\end{equation}
where $C_{v}$ is the type of node $v$'s attributes. $\mathbf{h}_{i}$ is the $i$-th attributes of node $v$. A bi-directional Long Short-Term Memory (Bi-LSTM) \cite{huangXY15} is used to fuse the embeddings learned by multiple attribute encoder $\mathcal{FC}$. Neighbor aggregation aims to aggregate the nodes with same type by using a Bi-LSTM to capture the position information:
\begin{equation}
	f_{2}^{t}(v)=\frac{\sum_{v^{'} \in N_{t}(v)}[\overrightarrow{LSTM}\{f_{1}(v^{'})\} \oplus \overleftarrow{LSTM}\{f_{1}(v^{'})\}]}{|N_{t}(v)|},
\end{equation}
where $N_{t}(v)$ is the first-order neighbors of node $v$ with type $t$. Type aggregation uses an attention mechanism to mix the embeddings of different types and produces the final node embeddings. 
\begin{equation}
	\mathbf{h}_{v}=\alpha^{v,v}f_{1}(v)+\sum_{t \in O_{v}}\alpha^{v,t}f_{2}^{t}(v).
\end{equation}
where $\mathbf{h}_{v}$ is the final embedding of node $v$. $O_{v}$ denotes the set of node types. Finally, a heterogeneous skip-gram loss is used as the unsupervised graph context loss to update the node embeddings. Through the three aggregation methods, HetGNN can preserve the heterogeneity of both graph structures and node attributes.

Other unsupervised methods either capture the heterogeneity of node attributes or the heterogeneity of graph structures. HNE \cite{chang2015heterogeneous} is proposed to learn embeddings for the cross-model data in HG, but it ignores the various types of links. SHNE \cite{zhang2019shne} focuses on capturing the semantic information of nodes by designing a deep semantic encoder with gated recurrent units (GRU) \cite{chung2014empirical}. Although it uses heterogeneous skip-gram to preserve the heterogeneity of graph, SHNE is designed specifically for text data. Cen \emph{et al.} propose GATNE \cite{cen2019representation}, which aims to learn node embeddings in multiplex graph, i.e., a heterogeneous graph with different types of edges. Compared with HetGNN, GATNE pays more attention to distinguishing different link relationships between the node pairs.


\subsubsection{Semi-supervised HGNNs}

Different from unsupervised HGNNs, semi-supervised HGNNs aim to learn task-specific node embeddings in an end-to-end manner. For this reason, they prefer to use attention mechanism to capture the most relevant structural and attribute information to the task.

\begin{figure}[ht]
	\center
	\includegraphics[width=\linewidth]{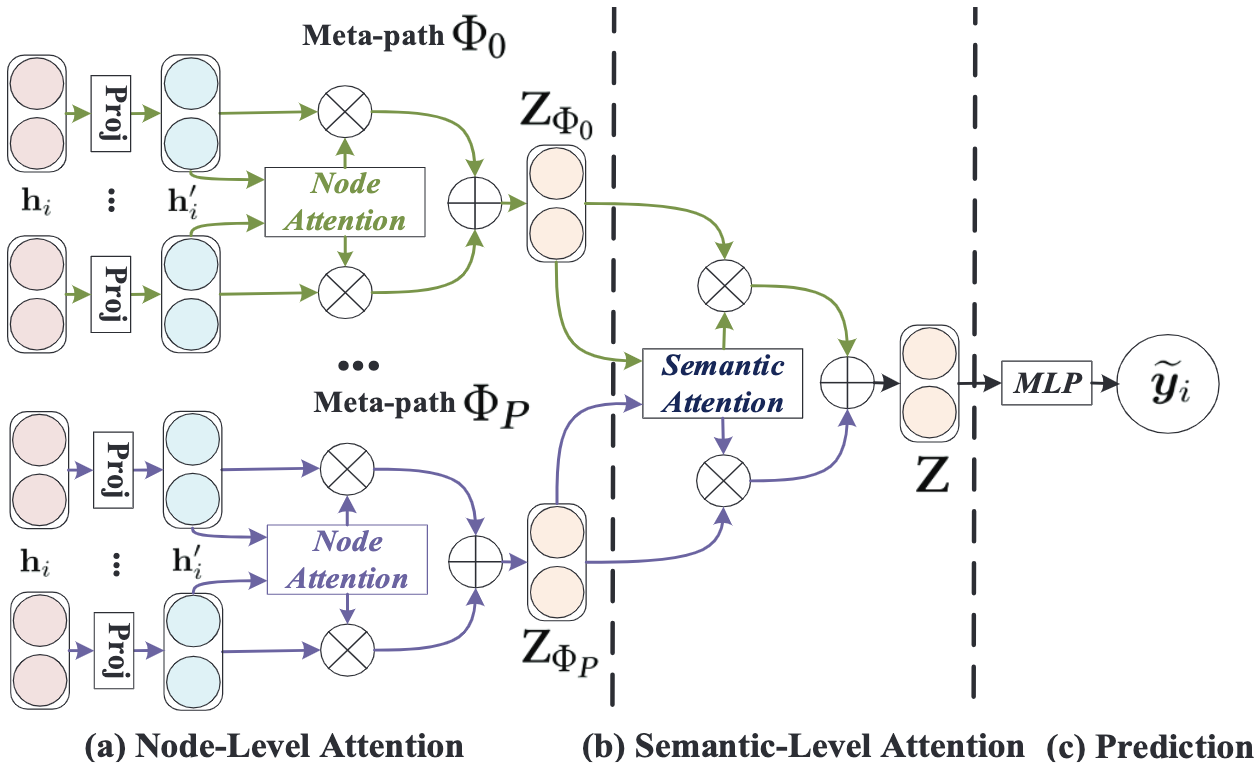}
	\caption{The architecture of HAN \cite{wang2019heterogeneous}. The whole model can be divided into three parts: Node-Level Attention aims to learn the importance of neighbors' features. Semantic-Level Attention aims to learn the importance of different meta-paths. Prediction layer utilizes the labeled nodes to update the node embeddings.}
	\label{HAN}
\end{figure}

Wang \emph{et al.} \cite{wang2019heterogeneous} propose heterogeneous graph attention network (HAN), which uses a hierarchical attention mechanism to capture both node and semantic importance. The architecture of HAN is shown in Fig. \ref{HAN}. 

It consists of three parts: node-level attention, semantic-level attention and prediction. Node-level attention aims to utilize self-attention mechanism \cite{vaswani2017attention} to learn the importances of neighbors in a certain meta-path:
\begin{equation}
	\alpha_{ij}^{m}=\frac{\exp(\sigma(\mathbf{a}_{m}^{\mathrm{T}} \cdot [\mathbf{h}_{i}^{'} \| \mathbf{h}_{j}^{'}]))}{\sum_{k \in \mathcal{N}_{i}^{m}} \exp(\sigma(\mathbf{a}_{m}^{\mathrm{T}} \cdot [\mathbf{h}_{i}^{'} \| \mathbf{h}_{k}^{'}]))},
\end{equation}
where $\mathcal{N}_{i}^{m}$ is the neighbors of node $i$ in meta-path $m$, $\alpha_{ij}^{m}$ is the weight of node $j$ to node $i$ under meta-path $m$. The node-level aggregation is defined as:
\begin{equation}
	\mathbf{h}_{i}^{m}=\sigma \left(\sum_{j \in \mathcal{N}_{i}^{m}}\alpha_{ij}^{m} \cdot \mathbf{h}_{j} \right),
\end{equation}
where $\mathbf{h}_{i}^{m}$ denotes the learned embedding of node $i$ based on meta-path $m$. Because different meta-paths capture different semantic information of HG, a semantic-level attention mechanism is designed to calculated the importance of meta-paths. Given a set of meta-paths $\{m_{0},m_{1}, \cdots, m_{P}\}$, after feeding node features into node-level attention, it has $P$ semantic-specific node embeddings $\{\mathbf{H}_{m_{0}},\mathbf{H}_{m_{1}}, \cdots, \mathbf{H}_{m_{P}}\}$. To effectively aggregate different semantic embeddings, HAN designs a semantic-level attention mechanism:
\begin{equation}
	w_{m_{i}}=\frac{1}{|\mathcal{V}|}\sum_{i \in \mathcal{V}}\mathbf{q}^{\mathrm{T}} \cdot \tanh(\mathbf{W} \cdot \mathbf{h}_{i}^{m} + \mathbf{b}),
\end{equation}
where $\mathbf{W} \in \mathbb{R}^{d' \times d}$ and $\mathbf{b} \in \mathbb{R}^{d' \times 1}$ denote the weight matrix and bias of the MLP, respectively. $\mathbf{q} \in \mathbb{R}^{d' \times 1}$ is the semantic-level attention vector. In order to prevent the node embeddings from being too large, HAN uses the softmax function to normalize $w_{m_{i}}$. Hence, the semantic-level aggregation is defined as:
\begin{equation}
	\mathbf{H}=\sum_{i=1}^{P}\beta_{m_{i}} \cdot \mathbf{H}_{m_{i}},
\end{equation}
where $\beta_{m_{i}}$ denotes the normalized $w_{m_{i}}$, which represents the semantic importance. $\mathbf{H} \in \mathbb{R}^{N \times d}$ denotes the final node embeddings. Finally, a task-specific layer is used to fine-tune the node embeddings with a small number of labels and the embeddings $\mathbf{H}$ can be used in the downstream tasks, such as node clustering and link prediction. HAN is the first to extend GNN to the heterogeneous graph and design a hierarchical attention mechanism, which can capture both structural and semantic information.

Subsequently, a series of attention-based HGNNs was proposed \cite{fu2020magnn, hong2020attention, fu2020magnn, HGT, encoder2019fu}. MAGNN \cite{fu2020magnn} designs intra-metapath aggregation and inter-metapath aggregation. The former samples some meta-path instances surrounding the target node and uses an attention layer to learn the importance of different instances, and the latter aims to learn the importance of different meta-paths. HetSANN \cite{hong2020attention} and HGT \cite{HGT} treat one type of node as query to calculate the importance of other types of nodes around it, through which the method can not only capture the interactions among different types of nodes, but also assign different weights to neighbors during aggregation. \cite{encoder2019fu} uses meta-paths as virtual edges to enhance the performance of graph attention operator.

In addition, there are some HGNNs that focus on other issues. NSHE \cite{NSHE} proposes to incorporate network schema, instead of meta-path, in aggregating neighborhood information. GTN \cite{GTN} aims to automatically identify the useful meta-paths and high-order links in the process of learning node embeddings. RSHN \cite{zhu2019RSHN} uses both original node graph and coarsened line graph to design a relation-structure aware HGNN. RGCN \cite{RGCN} uses multiple weight matrices to project the node embeddings into different relation spaces, thus capturing the heterogeneity of the graph.


\subsubsection{Summary}
As we can see that there are two ways to solve the heterogeneity of attributes: one is to use different encoders or type-specific transformation matrices to map the different attributes into a same space, such as \cite{zhang2019heterogeneous, chang2015heterogeneous}. Another is to treat meta-path as a special edge to connect the nodes with same type, such as \cite{wang2019heterogeneous, fu2020magnn}. Compared with shallow models, HGNNs have an obvious advantage that they have the ability of inductive learning, i.e., learning embeddings for the out-of-sample nodes \cite{hou2019alphacyber}. Besides, HGNNs need smaller memory space because they only need to store model parameters. These two reasons are important for the real-world applications. However, they still suffer from the huge time costing in inferencing and retraining.

\subsection{Application-oriented HG Embedding}
\label{application}
Heterogeneous graph embedding also has been closely combined with some specific applications, where the aforementioned information, e.g., attributes, is not sufficient for specific applications. Under such settings, one usually needs to carefully consider two factors: the first is how to construct a HG for a specific application, and the second is what information, i.e., domain knowledge, should be incorporated into heterogeneous graph embedding, so as to finally benefit the application.
In this section, we discuss three common types of applications: recommendation, identification and proximity search.

\subsubsection{Recommendation}
In recommendation system, the interaction among user and item can be naturally modeled as a HG with two types of nodes. Therefore, recommendation is a typical scenario that widely uses HG information \cite{shi2016survey}.
Besides, other types of information, such as the social relationships, can also be easily introduced in HG \cite{SemRec}, applying heterogeneous graph embedding to recommendation application is an important research field.

Early works recommend item to a user mainly based on meta-path aware similarity between user and item, such as HeteLearn \cite{HeteLearn} and SemRec \cite{SemRec}.
With the development of embedding technology, matrix factorization \cite{HeteroMF, FMG, HeteRec}, random walk \cite{shi2018heterogeneous} and advanced neural networks \cite{hu2018leveraging, han2018aspect, xu2019learning, wang2019unified, zhao2019intentgc, fan2019meta} are proposed to learn embeddings of user and item, so as to capture the complex interactions.

HERec \cite{shi2018heterogeneous} aims to learn the embeddings of users and items under different meta-paths and fuses them for recommendation.
It first finds the co-occurrence of users and items based on the meta-path guided random walks on user-item HG. 
Then it uses node2vec \cite{grover2016node2vec} to learn preliminary embeddings from the co-occurrence sequences of users and items. Because the embeddings under different meta-paths contain different semantic information, for better recommendation performance, HERec designs a fusion function to unify the multiple embeddings:
\begin{equation}
	g(\mathbf{h}_{u}^{m})=\frac{1}{|P|}\sum_{m=1}^{M}(\mathbf{W}^{m}\mathbf{h}_{u}^{m}+\mathbf{b}^{m}),
\end{equation}
where $\mathbf{h}_{u}^{m}$ is the embedding of user node $u$ in meta-path $m$. $M$ denotes the set of meta-paths. The fusion of item embeddings is similar to users. Finally, a prediction layer is used to predict the items that users prefer. HERec optimizes the graph embedding and recommendation objective jointly.

Apart from random walk, some methods try to use matrix factorization to learn user and item embeddings. HeteRec \cite{HeteRec} considers the implicit user feedback in HG. HeteroMF \cite{HeteroMF} designs a heterogenous matrix factorization technique to consider the context dependence of different types of nodes. FMG \cite{FMG} incorporates meta-graphs into embedding technology, which can capture some special patterns between users and items.

Previous methods mainly use shallow models to learn the embeddings of users and items, where the ability of express nonlinear interaction between them is limited. Therefore, some neural network-based methods are proposed.
One of the most important techniques is attention mechanism, which aims to find the important users and items in HG based recommendation.
MCRec \cite{hu2018leveraging} designs a neural co-attention mechanism to capture the relationship between user, item and meta-path. Specifically, it uses the users and items to find the important meta-paths. Meanwhile, the important meta-paths are used to find the important users and items in recommendation. Through this mutual selective attention mechanism, MCRec can not only learn embeddings of users, items and meta-paths, but also capture the complex interactions among them.
NeuACF \cite{han2018aspect} and HueRec \cite{wang2019unified} first calculate multiple meta-path-based commuting matrices, where each row represents the user-user similarity or item-item similarity. Then an attention mechanism is designed to learn the importance of different meta-path-based commuting matrices, so as to capture different semantic information.

Another type of important techniques is graph neural networks.
PGCN \cite{xu2019learning} converts the user-item interaction sequences into item-item graph, user-item graph and user-sequence graph. Then it designs a HGNN to propagate user and item information in the three graphs, so as to capture the collaborative filtering signals.
MEIRec \cite{fan2019meta} focuses on the problem of intent recommendation in E-commerce, which aims to automatically recommend user intent according to user historical behaviors. It constructs a user-item-query heterogeneous graph and designs a meta-path-guided HGNN to learn the embedding of users, items and queries, which can capture the intent of users.
GNewsRec \cite{hu2020graph} and GNUD \cite{GNUD} are designed for news recommendation. They consider both the content information of news and the collaborative information between users and news.
\cite{basket2020liu} employs graph convolutional network on heterogeneous graphs for basket recommendation.


\subsubsection{Identification}
Identification is to find the most likely nodes according to the given conditions on HG. For example, finding potential authors of a given paper or identifying users in cross-platform.
Currently, two representative identification applications, author identification \cite{chen2017task, zhang2018camel, park2019task} and user identification \cite{zhang2019key, fan2019idev, zhang2019your}, have been studied based on heterogeneous graph embedding.

Author identification aims to find the potential authors for an anonymous paper in the academic network. Camel \cite{zhang2018camel} aims to consider both the content information, e.g., the text of papers, and context information, e.g., the co-occurrence of paper and author. For content information, it designs a content encoder to learn embeddings from the abstract of paper and a metric-based loss function is used to learn the pair-wise relations between authors and papers:
\begin{equation}
	\mathcal{L}_{Metric}=\xi + \|f(\mathbf{h}_{v})-\mathbf{h}_{u}\|^{2} - \|f(\mathbf{h}_{v})-\mathbf{h}_{u^{'}}\|^{2},
\end{equation}
where $\xi$ is the margin, $f(\cdot)$ represents the content encoder and $\mathbf{h}_{v}$, $\mathbf{h}_{u}$ and $\mathbf{h}_{u^{'}}$ denote the attributes of paper, positive author and negative author, respectively.
For context information, a meta-path guided walk integrative learning module (MWIL) is proposed to preserve the graph structures:
\begin{equation}
	\mathcal{L}_{MWIL}=-\log \sigma[f(\mathbf{h}_{v}) \cdot \mathbf{h}_{u}] - \log \sigma[-f(\mathbf{h}_{v}) \cdot \mathbf{h}_{u^{'}}].
\end{equation}
It is worth noting that $\mathcal{L}_{MWIL}$ is a special skip-gram technique, which aims to preserve the proximity of positive author $u$ of paper $v$ within a walk length.
Through optimizing $\mathcal{L}_{Metric}$ and $\mathcal{L}_{MWIL}$ jointly, Camel considers both the heterogeneous graph structures and the pair-wise relation of author-paper.
Similar to the idea of Camel, PAHNE \cite{chen2017task} considers the pair-wise relations and TaPEm \cite{park2019task} maximizes the proximity between the paper-author pair and the context path around them.

Compared with author identification, user identification does not contain the pair-wise relation, i.e., user and paper. Therefore, it focuses on learning discriminating user embeddings with weak supervision information so that the target users can be identified more easily. Player2vec \cite{zhang2019key}, AHIN2vec \cite{fan2019idev} and Vendor2vec \cite{zhang2019your} are the principal methods.
They can be summarized as a general framework: first, some advanced neural networks, e.g., convolutional neural network (CNN) or recurrent neural network (RNN), are used to learn preliminary node embeddings from the raw features. Then the preliminary node embeddings will be propagated on the graphs, constructed by different meta-paths, to utilize the neighborhood information.
Finally, a semi-supervised loss function is used to make the node embeddings contain application-specific information. Under the guidance of partially labeled nodes, the node embeddings can distinguish special users from the ordinary users in the graph, which can be used for user identification.

\subsubsection{Proximity Search}
\begin{figure}[ht]
\center
\includegraphics[width=\linewidth]{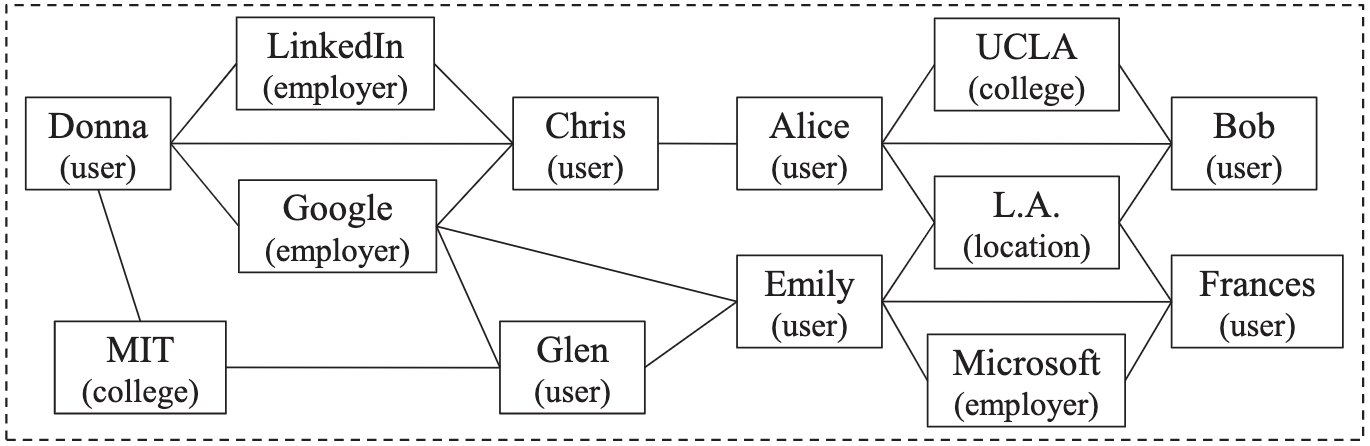}
\caption{An example of semantic proximity search \cite{liu2018interactive}, which gives a query object (e.g., \emph{Alice}) and requires the method rank other objects according to the semantic relation (e.g., "who are likely to be her \emph{schoolmates}?").}
\label{proximity}
\end{figure}

Given a target node in HG, the proximity search, as shown in Fig. \ref{proximity}, is to find the nodes that are closest to the target node by using structural and semantic information of HG. Some earlier studies have deal with this problem in homogeneous graphs, for example, web search \cite{jeh2003scaling}. Recently, some methods try to utilize HG in proximity search \cite{sun2011pathsim, shi2014hetesim}. However, these methods only use some statistical information, e.g., the number of connected meta-paths, to measure the similarity of two nodes in HG, which lack flexibility. With the development of deep learning, some embedding methods are proposed.

Prox \cite{liu2017semantic} uses heterogeneous graph embedding in semantic proximity search. Given a set of training tuples $\{q_i, v_i, u_i\}$, where $q_i$ is the query node and in each query the similarity $S(q_i, v_i)$ between node $v_i$ and $q_i$ is larger than $S(q_i, u_i)$.
It firstly samples some heterogeneous sequences for each node in the training tuples and feed them into a LSTM ito learn node embeddings. A ranking-based loss function is used to use the implicit supervision information:
\begin{equation}
	L(S(q_i, v_i),S(q_i, u_i))=-\log \sigma(S(q_i, v_i) - S(q_i, u_i)).
\end{equation}
Minimizing the function indicates that the similarity between $v_i$ and $q_i$ should be larger than that between $u_i$ and $q_i$. Different from previous methods that use manually calculated similarities of nodes in HG, Prox uses heterogeneous graph embedding to avoid the feature engineering for semantic proximity search, which is an efficient and effective approach.

Then a series of methods are proposed. IPE \cite{liu2018interactive} considers the interactions among different meta-path instances and propose an interactive-paths structure to improve the performance of heterogeneous graph embedding. SPE \cite{liu2018subgraph} proposes a subgraph-augmented heterogeneous graph embedding method, which uses a stacked autoencoder to learn the subgraph embedding so as to enhance the effect of semantic proximity search. D2AGE \cite{liu2018distance} explores the DAG structure for better measuring the similarity between two nodes and designs a DAG-LSTM to learn node embeddings. 

\subsubsection{Summary}
Incorporating heterogeneous graph embedding into specific applications usually need to consider the domain knowledge. For example, in recommendation, meta-path ``user-item-user'' can be used to capture the user-based collaborative filtering, while ``item-user-item'' represents the item-based collaborative filtering; in proximity search, methods use meta-paths to capture the semantic relationships between nodes, thus enhancing the performance. Therefore, utilizing HG to capture the application-specific domain knowledge is essential for application-oriented heterogeneous graph embedding.

\subsection{Dynamic HG Embedding}
In the beginning of Section \ref{taxonomy}, we mention that previous HG surveys \cite{dong2020survey, yang2020survey} focus on summarizing the static methods, while the dynamic methods are largely ignored. Since the real-world graphs are constantly changing over time, to fill this gap, in this section, we summary the dynamic heterogeneous graph embedding methods. Specifically, they can be divided into two categories: incremental update and retrained update methods. The former learns the embedding of new node in the next timestamp by utilize existing node embeddings, while the latter will retrain the models in each timestamp. Both of them have its own pros and cons, and will be discussed in the end.

\subsubsection{Incremental HG Embedding}

DyHNE \cite{DyHNE} is an incremental update method based on the theory of matrix perturbation, which learns node embeddings while considering both the heterogeneity and evolution of HG. To ensure the effectiveness, DyHNE preserves the meta-path based first- and second-order proximities. The first-order proximity
requires two nodes connected by meta-path $m$ to have similar embeddings. And the second-order proximity
indicates that the node embedding should be close to the weighted sum of its neighbor embeddings. Specifically, the first- and second-order proximities can be uniformly rewritten as:
\begin{equation}
	\mathcal{L}=\operatorname{tr}(\mathbf{H}^{\top}(\mathbf{L}+\gamma \mathbf{T})\mathbf{H}),
\end{equation}
where $\gamma$ is a hyperparameter. $\mathbf{W}=\sum_{m \in M}\theta_{m}\mathbf{W}^{m}$ and $\mathbf{D}=\sum_{m \in M}\theta_{m}\mathbf{D}^{m}$ are the fusion of different meta-paths, which lead to $\mathbf{L}=\mathbf{D}-\mathbf{W}$ and $\mathbf{T}=(\mathbf{I}-\mathbf{W})^{\top}(\mathbf{I}-\mathbf{W})$. The minimization of $\mathcal{L}$ can be solved by the eigenvalue decomposition:
\begin{equation}
	(\mathbf{L}+\gamma \mathbf{T})\mathbf{H}=\mathbf{D}\Lambda\mathbf{H},
\end{equation}
where $\Lambda=diag(\lambda_{1},\lambda_{2}\cdots\lambda_{N})$ is the eigenvalue matrix. To model the evolution of HG, DyHNE uses the perturbation of meta-path augmented adjacency matrices to naturally capture changes of graph. At a new timestamp, the matrix becomes:
\begin{gather}
	(\mathbf{L}+\Delta\mathbf{L} + \gamma\mathbf{T} + \gamma\Delta\mathbf{T})(\mathbf{h}_{i}+\Delta\mathbf{h}_{i}) \notag \\
	=(\lambda_{i}+\Delta\lambda_{i} )(\Lambda+\Delta\Lambda)(\mathbf{h}_{i}+\Delta\mathbf{h}_{i}).
\end{gather}
where $\Delta$ denote the perturbation term. $\Delta\mathbf{h}$ and $\Delta\lambda$ are the changes of the eigenvectors and eigenvalues.
Hence, the incremental update of node $i$ is  how to calculate the changes of the $i$-th eigen-pair $(\Delta\mathbf{h}_{i}, \Delta\lambda_{i})$.
With some approximations, DyHNE can directly update the node embeddings without retraining the whole model. 
Generally speaking, DyHNE preserves both the structural and semantic information of HG and uses the perturbation of matrix to capture the evolution of HG over time, which is an effective and efficient method.
\cite{change2vec, MetaDynaMix} also adopt the idea of incremental update. Change2vec \cite{change2vec} proposes a dynamic version of metapath2vec. MetaDynaMix \cite{MetaDynaMix} uses the incremental update on the matrix factorization of HG. 

\subsubsection{Retrained HG Embedding}

Retrained update methods first use GNNs to learn node or edge embeddings in each timestamp and then design some advanced neural network, e.g., RNN or attention mechanism, to capture the temporal information of HG.

\begin{figure}[ht]
	\center
	\includegraphics[width=\linewidth]{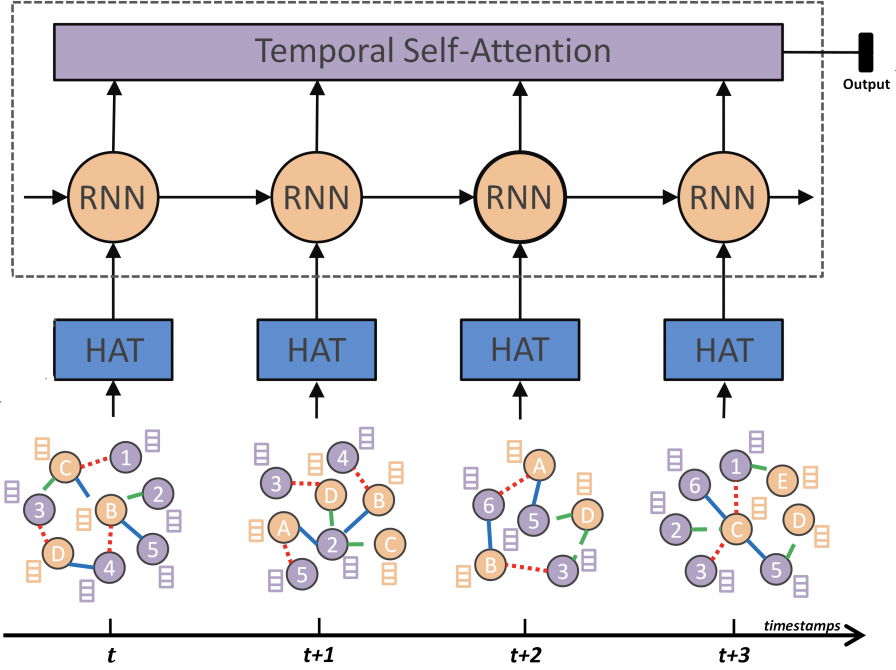}
	\caption{The architecture of DyHATR \cite{DyHATR}. It consists of two parts: first, a hierarchical attention mechanism is designed to learn node embeddings by fusing the attributes of neighbors. Then, a RNN with self-attention mechanism is used to capture the temporal information.}
	\label{fig_DyHATR}
\end{figure}
 
DyHATR \cite{DyHATR} aims to capture the temporal information through the changes of nodes embeddings in different timestamps. To this end, as shown in Fig. \ref{fig_DyHATR}, it first designs a hierarchical attention mechanism (HAT), which contains a node- and edge-level attention, to learn node embeddings by fusing the attributes of neighbors. The node-level attention is defined as:
\begin{equation}
	\alpha^{rt}_{i,j}=\frac{\exp(\sigma(\mathbf{a}^\mathrm{T}_{r} \cdot [\mathbf{M}_{r} \cdot \mathbf{h}_{i} || \mathbf{M}_{r} \cdot \mathbf{h}_{j}]))}{\sum_{k \in \mathcal{N}_{i}^{rt}} \exp(\sigma(\mathbf{a}^\mathrm{T}_{r} \cdot [\mathbf{M}_{r} \cdot \mathbf{h}_{i} || \mathbf{M}_{r} \cdot \mathbf{h}_{j}]))},
\end{equation}
where $\mathcal{N}_{i}^{rt}$ represents the neighbors of node $i$ in edge type $r$ and timestamp $t$, and $\mathbf{a}_{r}$ is the attention vector. And the edge-level attention is:
\begin{equation}
	\beta^{rt}_{i}=\frac{\exp (\mathbf{q}^{\mathrm{T}} \cdot \sigma(\mathbf{W} \cdot \mathbf{h}_{i}^{rt} + \mathbf{b}))}{\sum_{r \in R} \exp (\mathbf{q}^\mathrm{T} \cdot \sigma(\mathbf{W} \cdot \mathbf{h}_{i}^{rt} + \mathbf{b}))},
\end{equation}
where $\mathbf{q}^\mathrm{T}$ is the attention vector in edge-level attention. Through the node- and edge-level attention, DyHATR can learn the node embeddings under different timestamps. In order to capture the temporal information hidden in the changes of node embeddings, the node embeddings are fed into a RNN in the order of timestamps. Coincidentally, DyHAN \cite{DyHAN} also designs a hierarchical attention mechanism to learn the importance of nodes and timestamps, respectively.

\subsubsection{Summary}

It can be seen that the incremental update methods are efficient, but they can only capture the short-term temporal information (i.e., the last timestamp) \cite{DyHATR}. Besides, incremental update methods focus on utilizing shallow model, which lacks the non-linear expressive power. On the contrary, the retrained update methods employ neural networks to capture the long-term temporal information. However, they suffer from the high computational cost. Therefore, how to combine the advantages of these two kinds of models is an important problem. In addition, there are some meaningful problems to consider, e.g., how to eliminate the cumulative errors in incremental update methods.

\subsection{Miscellanea}
In the previous section, we introduce the major applications in heterogeneous graph embedding. There are also some other methods that do not belong to the existing categories. Here, we briefly introduce them.

The first is incorporating HG into natural language processing (NLP). Due to the multiple elements in the corpus, e.g., words, entities, sentences or paragraphs, many NLP tasks can be modeled as HG naturally. Graph-to-sequence (Graph2Seq) learning is an important topic in NLP, which aims to transform the graph-structured embeddings to word sequences for text generation \cite{g2s2018amr, g2s2018ggnn}. AMR-to-text generation is a typical Graph2Seq task. It generates text from the Abstract Meaning Representation (AMR) graph, where nodes represent the semantic concepts in the text and edges denote the relations between concepts. In order to learn useful information from the AMR graph, Yao \emph{et al.} \cite{g2s2020hg} treat AMR graph as a heterogeneous graph and design a heterogeneous graph encoder to learn the semantic information among the concepts. Besides, Hu \emph{et al.} \cite{HGAT} propose HGAT for short text classification, which treats the topic, entities and documents as a HG and designs a hierachical attention mechanism to learn the similarity among short texts. GNewsRec \cite{hu2020graph} and GNUD \cite{GNUD} use HG to model the collaborative filtering between news and users in news recommendation task. \cite{topicmodel} incorporates HG into topic model for aspect mining. \cite{fake2019zhang} uses HG in fake new detection.

Similar to NLP, multi-modal data can also be modeled by HG due to the various data forms, e.g., text, images or videos. The potential dependencies and connections among multi-model data can be modeled by HG easily. Therefore, some methods try to use heterogeneous graph embedding to capture the potential dependencies and connections. For example, Community Question Answering (CQA) aims to recommend the suitable answers for each question. Because the answers and questions may contain text and pictures, \cite{CQA1, CQA2} treats the answers and question as a heterogeneous graph to capture the potential connections, thus making the state-of-the-art performance.

Besides, graph embedding in hyperbolic space has received widespread attention \cite{hyperbolic, hyperGNN1, hyperGNN2}. Because whether Euclidean spaces are the optimal isometric spaces is still an unsolved problem, exploring heterogeneous graph embedding in the hyperbolic spaces is a meaningful research direction. \cite{wang2019hyperbolic} shows that hyperbolic spaces can capture the hierarchical and power-law structure of the heterogeneous graph, which provides a theoretical guarantee for the future work to some extent.

Moreover, HG embedding are also widely used to model many other tasks, such as entity set expansion \cite{shi2021entity}, basket recommendation \cite{liu2020basconv}, event categorization \cite{peng2019event} and social network \cite{zhang2017blmne}.

\section{Technique Summary}
\label{model}

\begin{table*}[t]
  \centering
  \setlength{\belowcaptionskip}{0.05cm}
  \renewcommand\arraystretch{1.25}
  \caption{Typical heterogeneous graph embedding methods.}
    \resizebox{\linewidth}{!}{
    \begin{tabular}{|c|c|c|c|c|c|l|}
    \hline
    Method & \multicolumn{1}{c|}{Inductive} & \multicolumn{1}{c|}{\quad Label \quad\quad} & Information & Task  & Technique & \multicolumn{1}{c|}{Characteristic} \\
    \hline
    \hline
    mp2vec \cite{dong2017metapath2vec} &       &       & \multirow{6}[0]{*}{Strcuture} & \multirow{6}[0]{*}{Embedding} & \multirow{7}[0]{*}{\tabincell{c}{Random walk\\(Shallow model)}} & \multirow{7}[0]{*}{\tabincell{l}{\tabitem Easy to parallelize \\ \tabitem Two-stage training \\ \tabitem High memory cost  \\ \\ Complexity: $\mathcal{O}(\tau \cdot l \cdot k \cdot n_{s} \cdot d \cdot |\mathcal{V}|)$} } \\
\cline{1-3}    Spacey \cite{he2019hetespaceywalk} &       &       &       &       &       &  \\
\cline{1-3}    JUST \cite{hussein2018meta}  &       &       &       &       &       &  \\
\cline{1-3}    BHIN2vec \cite{lee2019bhin2vec} &       &       &       &       &       &  \\
\cline{1-3}    HHNE \cite{wang2019hyperbolic} &       &       &       &       &       &  \\
\cline{1-3}    mg2vec \cite{zhang2018metagraph2vec} &       &       &       &       &       &  \\
\cline{1-5}    HeRec \cite{shi2018heterogeneous} &       & $\surd$ & Strcuture+Task & Recommendation &       &  \\
    \hline
    PME \cite{chen2018pme}  &       &       & \multirow{6}[0]{*}{Strcuture} & \multirow{15}[0]{*}{Embedding} & \multirow{6}[0]{*}{\tabincell{c}{Decomposition\\(Shallow model)} } & \multirow{6}[0]{*}{\tabincell{l}{\tabitem Easy to parallelize \\ \tabitem Two-stage training \\ \tabitem High memory cost \\ \\ Complexity: $\mathcal{O}(|\mathcal{E}| \cdot d)$} } \\
\cline{1-3}    EOE \cite{xu2017embedding}  &       &       &       &       &       &  \\
\cline{1-3}    HEER \cite{shi2018easing} &       &       &       &       &       &  \\
\cline{1-3}    MNE \cite{zhang2018scalable}  &       &       &       &       &       &  \\
\cline{1-3}    PTE \cite{chen2018pme}  &       &       &       &       &       &  \\
\cline{1-3}    RHINE \cite{lu2019relation} &       &       &       &       &       &  \\
\cline{1-4}\cline{6-7}    HAN \cite{wang2019heterogeneous}  & $\surd$ & $\surd$ & \multirow{9}[0]{*}{Structure+Attribute} &       & \multirow{15}[0]{*}{\tabincell{c}{Message passing\\(Deep model)} } & \multirow{15}[0]{*}{\tabincell{l}{\tabitem End-to-End training \\ \tabitem Encoding structures and attributes  \\ \tabitem Semantic fusion \\ \tabitem High training cost \\ \\Complexity: $\mathcal{O}(|\mathcal{V}| \cdot d_{1} + |\mathcal{R}| \cdot d_{2})$} } \\
\cline{1-3}    MAGNN \cite{fu2020magnn} & $\surd$ & $\surd$ &       &       &       &  \\
\cline{1-3}    HetSANN \cite{hong2020attention} & $\surd$ & $\surd$ &       &       &       &  \\
\cline{1-3}    HGT \cite{HGT}  & $\surd$ & $\surd$ &       &       &       &  \\
\cline{1-3}    HetGNN \cite{zhang2019heterogeneous} & $\surd$ &       &       &       &       &  \\
\cline{1-3}    GATNE \cite{cen2019representation} & $\surd$ &       &       &       &       &  \\
\cline{1-3}    GTN \cite{GTN}  &       & $\surd$ &       &       &       &  \\
\cline{1-3}    RSHN \cite{zhu2019RSHN} &       & $\surd$ &       &       &       &  \\
\cline{1-3}    RGCN \cite{RGCN} & $\surd$ & $\surd$ &       &       &       &  \\
\cline{1-5}    IntentGC \cite{zhao2019intentgc} & $\surd$ & $\surd$ & \multirow{6}[0]{*}{\tabincell{c}{Strcuture\\+Attribute+Task} } & \multirow{3}[0]{*}{Recommendation} &       &  \\
\cline{1-3}    MEIRec \cite{fan2019meta} & $\surd$ & $\surd$ &       &       &       &  \\
\cline{1-3}    GNUD \cite{GNUD} & $\surd$ & $\surd$ &       &       &       &  \\
\cline{1-3}\cline{5-5}    Player2vec \cite{zhang2019key} & $\surd$ & $\surd$ &       & \multirow{3}[0]{*}{Identification} &       &  \\
\cline{1-3}    AHIN2vec \cite{fan2019idev} & $\surd$ & $\surd$ &       &       &       &  \\
\cline{1-3}    Vendor2vec \cite{zhang2019your} & $\surd$ & $\surd$ &       &       &       &  \\
    \hline
    HIN2vec \cite{fu2017hin2vec} &       &       & \multirow{2}[0]{*}{Strcuture} & \multirow{5}[0]{*}{Embedding} & \multirow{8}[0]{*}{\tabincell{c}{Encoder-decoder\\(Deep model)} } & \multirow{8}[0]{*}{\tabincell{l}{\tabitem End-to-End training \\ \tabitem Flexible goal-orientation \\ \\Complexity: $\mathcal{O}(|\mathcal{V}| \cdot d_{1} + |\mathcal{E}| \cdot d_{2})$} } \\
\cline{1-3}    DHNE \cite{tu2018structural} &       &       &       &       &       &  \\
\cline{1-4}    HNE \cite{chang2015heterogeneous}  & $\surd$ & $\surd$ & \multirow{3}[0]{*}{Structure+Attribute} &       &       &  \\
\cline{1-3}    SHNE \cite{zhang2019shne} &       & $\surd$ &       &       &       &  \\
\cline{1-3}    NSHE \cite{NSHE} &       &       &       &       &       &  \\
\cline{1-5}    PAHNE \cite{chen2017task} &       & $\surd$ & \multirow{3}[1]{*}{\tabincell{c}{Strcuture\\+Attribute+Task} } & \multirow{3}[1]{*}{Identification} &       &  \\
\cline{1-3}    Camel \cite{zhang2018camel} &       & $\surd$ &       &       &       &  \\
\cline{1-3}    TaPEm \cite{park2019task} &       & $\surd$ &       &       &       &  \\
    \hline
    HeGAN \cite{hu2019adversarial} &       &       & \multirow{2}[0]{*}{Strcuture} & \multirow{2}[0]{*}{Embedding} & \multirow{3}[0]{*}{\tabincell{c}{Adversarial\\(Deep model)} } & \multirow{3}[0]{*}{\tabincell{l}{\tabitem Robustness \\ \tabitem High complexity \\Complexity: $\mathcal{O}(|\mathcal{V}| \cdot |\mathcal{R}| \cdot n_{s} \cdot d)$} } \\
\cline{1-3}    MV-ACM \cite{MV-ACM} &       &       &       &       &       &  \\
\cline{1-5}    Rad-HGC \cite{hou2019alphacyber} &       & $\surd$ & Strcuture+Task & Malware detection &       &  \\
    \hline
    \end{tabular}}
  \label{typicalworks}
\end{table*}

In the previous section, we category the heterogeneous graph embedding methods based on different problem setting.
In this section, from the technical perspective, we summarize the widely used techniques (or models) in heterogeneous graph embedding, which can be generally divided into two categories: shallow model and deep model.

\subsection{Shallow Model}
Early heterogeneous graph embedding methods focus on employing shallow model. They first initialize the node embeddings randomly, and then learn the node embeddings through optimizing some well-designed objective functions.
We divide the shallow model into two categories: random walk-based and decomposition-based.

\textbf{Random walk-based.} In homogeneous graph, random walk, which generates some node sequences in a graph, is usually used to capture the local structure of a graph \cite{grover2016node2vec}. While in heterogeneous graph, the node sequence should contain not only the structural information, but also the semantic information. Therefore, a series of semantic-aware random walk techniques are proposed \cite{zhang2018scalable, dong2017metapath2vec, he2019hetespaceywalk, hussein2018meta, lee2019bhin2vec, wang2019hyperbolic, shi2018heterogeneous}. For example, metapath2vec \cite{dong2017metapath2vec} uses meta-path-guided random walk to capture the semantic information of two nodes, e.g., the co-author relationship in academic graph. Spacey \cite{he2019hetespaceywalk} and metagraph2vec \cite{zhang2018metagraph2vec} design metagraph-guided random walks, which preserve a more complex similarity between two nodes.

\textbf{Decomposition-based.} Decomposition-based techniques aim to decompose HG into several sub-graphs and preserve the proximity of nodes in each sub-graph \cite{chen2018pme, xu2017embedding, shi2018aspem, shi2018easing, matsuno2018mell, tang2015pte, gui2016large}. PME \cite{chen2018pme} decomposes the heterogeneous graph into some bipartite graphs according to the types of links and projects each bipartite graph into a relation-specific semantic space. PTE \cite{tang2015pte} divides the documents into word-word graph, word-document graph and word-label graph. Then it uses LINE \cite{tang2015line} to learn the shared node embeddings for each sub-graph. HEBE \cite{gui2016large} samples a series of subgraphs from a HG and preserves the proximity between the center node and its subgraph.

\subsection{Deep Model}
Deep model aims to use advanced neural networks to learn embedding from the node attributes or the interactions among nodes, which can be roughly divided into three categories: message passing-based, encoder-decoder-based and adversarial-based.

\textbf{Message passing-based.} The idea of message passing is to send the node embedding to its neighbors, which is always used in GNNs. The key component of message passing-based techniques is to design a suitable aggregation function, which can capture the semantic information of HG \cite{wang2019heterogeneous, fu2020magnn, hong2020attention, zhang2019heterogeneous, cen2019representation, NSHE, GTN, zhu2019RSHN, RGCN}. 
HAN \cite{wang2019heterogeneous} designs a hierarchical attention mechanism to learn the importance of different nodes and meta-paths, which captures both structural information and semantic information of HG.
HetGNN \cite{zhang2019heterogeneous} uses bi-LSTM to aggregate the embedding of neighbors so as to learn the deep interactions among heterogeneous nodes.
GTN \cite{GTN} designs an aggregation function, which can find the suitable meta-paths automatically during the process of message passing.

\textbf{Encoder-decoder-based.} Encoder-decoder-based techniques aim to employ some neural networks as encoder to learn embedding from node attributes and design a decoder to preserve some properties of the graphs \cite{tu2018structural, chang2015heterogeneous, zhang2019shne, chen2017task, zhang2018camel, park2019task}.
For example, HNE \cite{chang2015heterogeneous} focuses on multi-modal heterogeneous graph. It uses CNN and autoencoder to learn embedding from images and texts, respectively. Then it uses the embedding to predict whether there is a link between the images and texts. 
Camel \cite{zhang2018camel} uses GRU as encoder to learn paper embedding from the abstracts. A skip-gram objective function is used to preserve the local structures of the graphs.
DHNE \cite{tu2018structural} uses autoencoder to learn embedding for the nodes in a hyperedge. Then it designs a binary classification loss to preserve the indecomposability of the hyper-graph.

\textbf{Adversarial-based.} Adversarial-based techniques utilize the game between generator and discriminator to learn robust node embedding. In homogeneous graph, the adversarial-based techniques only consider the structural information, for example, GraphGAN \cite{wang2018graphgan} uses Breadth First Search when generating virtual nodes. In a heterogeneous graph, the discriminator and generator are designed to be relation-aware, which captures the rich semantics on HGs. HeGAN \cite{hu2018leveraging} is the first to use GAN in heterogeneous graph embedding. It incorporates the multiple relations into the generator and discriminator, so that the heterogeneity of a given graph can be considered. MV-ACM \cite{MV-ACM} uses GAN to generate the complementary views by computing the similarity of nodes in different views.

\subsection{Review}
In Table \ref{typicalworks}, we categorize the typical heterogeneous graph embedding methods through different perspectives. Specifically, from the left to right, we gradually coarsen the properties of each method, so as to summarize their commonalities.
 
The first two columns indicate whether the method has inductive capability and whether it needs labels for training. We can see that most message passing-based methods have the inductive capability because they can update the node embeddings by aggregating neighborhood information. But they need additional labels to guide the training process.

The middle two columns show the information and task in each method. It can be seen that most deep learning-based methods are proposed for HG with attributes or specific application, while the shallow model-based methods are mainly designed for the use of structures.
One possible reason is that HG with attributes or specific applications usually needs to introduce additional information or domain knowledge. However, modeling the domain knowledge may be complicated, and the relationship with HG may also need to be described carefully. Deep model provides a more powerful support for this kind of complex modeling, and it helps to make better progress in the complex application scenarios. Meanwhile, the emerging HGNNs can naturally integrate graph structures and attributes, so it is more suitable for the complex scenes and content.

The last two columns summarize the techniques used in HG embedding and their characteristics. Shallow models are easy to parallel. But they are two-stage training, i.e., the embeddings are not relevant to the downstream tasks, and the memory cost is heavy. On the contrary, deep models are end-to-end training and require less memory space. Besides, message passing-based techniques are good at encoding structures and attributes simultaneously, and integrating different semantic information. Compared with message passing-based techniques, encoder-decoder-based techniques are weak in fusing information due to the lack of messaging mechanism. But they are more flexible to introduce different objective function through different decoders. Adversarial-based methods prefer to utilize the negative samples to enhance the robustness of the embeddings. But the choice of negative samples has a huge influence on the performance, thus leading higher variances \cite{hu2019adversarial}.

It is worth noting that we also list the complexity of each techniques, where $\tau$ is the number of random walks, $l$ is the length of random walk, $k$ is the windows size in skip-gram \cite{mikolov2013distributed} and $n_{s}$ is the number of samples. The complexity of random walk technique consists of two parts: random walk and skip-gram, both of which are linear with the number of nodes. Decomposition technique needs to divide HGs into sub-graphs according to the type of edges, so the complexity is linear with the number of edges, which is higher than random walk. Message passing technique mainly uses node-level and semantic-level attention to learn node embeddings, so its complexity is related to the number of nodes and node types. As for the encoder-decoder technique, the complexity of encoder is related to the number of nodes, while decoder is usually used to preserve the network structures, so it is linear with the number of edges. Adversarial technique needs to generate the negative samples for each node, so the complexity is related to the number of nodes and negative samples.


\section{Real-world Deployed Systems}

Heterogeneous graph embedding is closely related with the real-world applications, as heterogeneous objects and interactions are ubiquitous in many practical systems. Here we focus on summarizing the industrial level applications with heterogeneous graph embedding. Different from those methods with specific tasks mentioned in Section~\ref{application}, methods introduced in this section solve practical problems in applications with industrial data. In addition, for industrial-level applications, we pay more attention to two key components: HG construction with industrial data and graph embedding techniques on the HG.

\begin{figure*}[htbp]
	\centering
	\subfigure[E-commerce recommendation HG \cite{zhao2019intentgc}.]{
		\includegraphics[height=1.2in,width=0.3\linewidth]{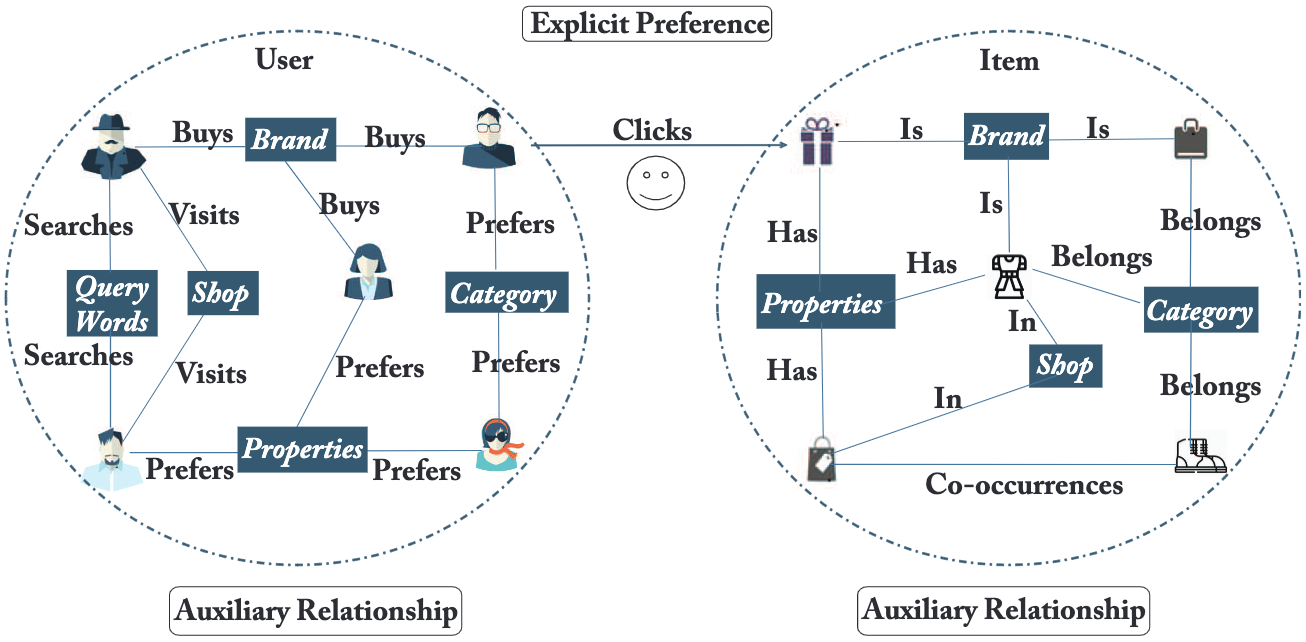}
		\label{fig:E-commerce HIN}
	}
	\hspace{7mm}
	\subfigure[Intent recommendation HG \cite{fan2019meta}.]{
		\includegraphics[height=1.5in,width=0.25\linewidth]{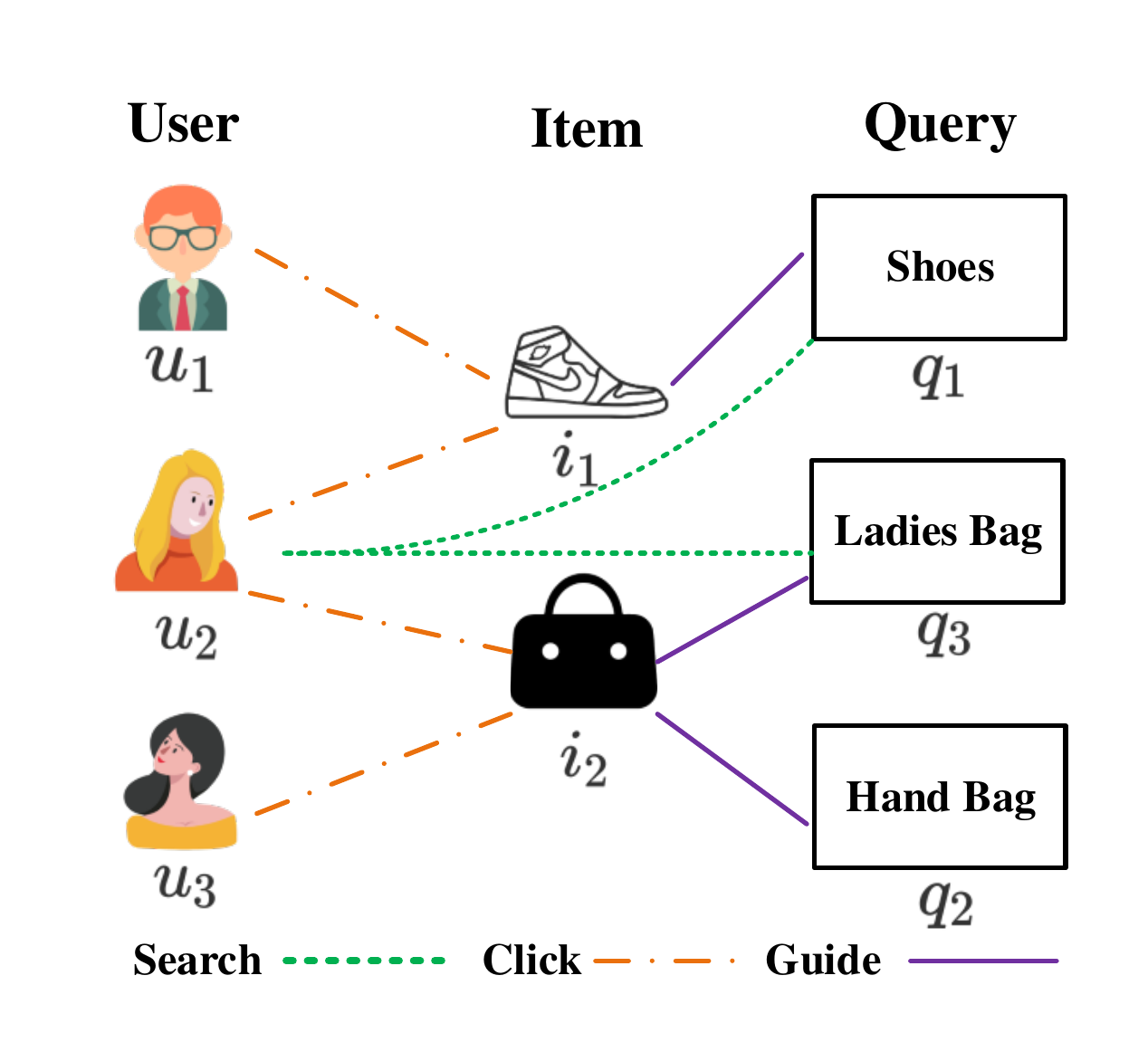}
		\label{fig:intent}
		
	}
	\hspace{7mm}
	\subfigure[User Profiling HG \cite{chen2019semi}.]{
		\includegraphics[height=1.4in,width=0.3\linewidth]{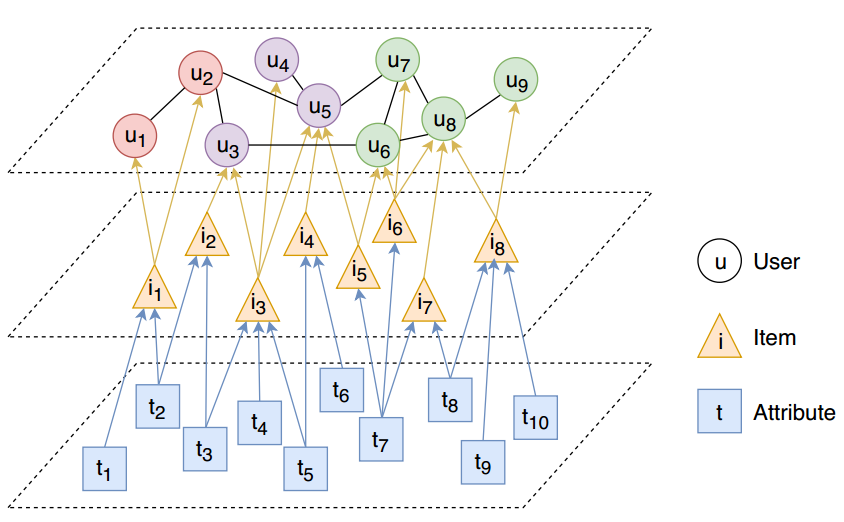}
		\label{fig:profiling}
	}
	\caption{The representative HGs in E-commerce.}
	\label{fig:ana}
\end{figure*}

\subsection{E-commerce}
\par E-commerce, such as Taobao\footnote{www.taobao.com} and Amazon\footnote{www.amazon.com}, is the activity of electronic trading of products on online services. It plays an important role in social economy development. Usually, large-scale heterogeneous objects and interactions, such as users, items, and shops, are involved in an e-commerce platform. Therefore, HG is a powerful and nature network analysis paradigm to model such complex data. HG embedding has been applied to various important services and tasks in e-commerce, such as item recommendation, intent recommendation, user profiling, and fraudster detection.

\par Recommendation is an important service of an e-commerce platform. A simple recommendation scenario primarily considers the interactions of users and items. However, due to the real business demands in e-commerce, it is highly desirable to comprehensively model users and items. HG can be used to model the interactions among users, items, and their auxiliary information~ \cite{SemRec}. As shown in Fig.~\ref{fig:E-commerce HIN}, the HG constructed by IntentGC~ \cite{zhao2019intentgc} is composed of user part and item part, and each part models the corresponding heterogeneous relationships. IntentGC translates the original HG as a multi-relation graph of users and items and develops a multi-relation graph convolution method to learn node embeddings. Besides integrating
the separate auxiliary information of user and item parts, GATNE~ \cite{cen2019representation} distinguishes the interactions between user and item pairs as multiple types, models this scenario as an attributed multiplex heterogeneous graph and proposes an unified embedding method that captures both attribute and edge information. More recently, to solve the interaction sparsity problem, Xu~\emph{et al.} \cite{xu2020gemini} transform the original user-item heterogeneous graph into two semi homogeneous graphs from the perspective of users and items respectively.

\par Different from recommending items for users, intent recommendation is a new type of recommendation service in mobile e-commerce Apps, which aims to automatically recommend user intent according to user historical behaviors without any input. Fan~\emph{et al.}~ \cite{fan2019meta} propose to represent user intent as default queries in search box and transform the intent recommendation problem as recommending the queries. They construct a HG containing three types of nodes (Users, Items and Queries) and their mutual interactions, shown in Fig. \ref{fig:intent}. Then, a meta-path-guided HGNN, called MEIRec, is designed to learn the nodes' embeddings of users and queries through aggregating the neighbors along the given meta-paths in an end-to-end manner.  

\par User profiling is playing an increasingly important role in providing personalized services in e-commerce platform. Different from previous methods only considering each user as an individual data instance, recent literature begins to model the abundant interaction information of users as a HG to enrich the characteristics of users. Chen \emph{et al.}~ \cite{chen2019semi} construct three kinds of objects (i.e., users, items and attributes) as a HG, shown in Fig.~\ref{fig:profiling}, and propose a hierarchical heterogeneous GAT to predict the traits of users (e.g., gender and age) by aggregating each layer of objects' embeddings. Apart from trait prediction, Zheng \emph{et al.}~ \cite{zheng2018heterogeneous} exploit HG to model the interactions between PID and MID with item ID in the e-commerce user alignment task. Then a Heterogeneous Embedding Propagation (HEP) model, encoding the interaction and edge features into node embeddings, is proposed to predict whether PID and MID across different devices refer to the same person.

\par With the development of e-commerce, there are many fraudsters in e-commerce system, who profit from transactions by illegal means. Due to the heterogeneity of fraudsters behavior patterns, some works try to detect these malicious accounts through HG embedding methods. Liu~\emph{et al.}~ \cite{GEM} consider behaviours of fraudsters as ``Device aggregation" and ``Activity aggregation" in the view of HG, and they propose a GNN, called GEM, which simultaneously models the topology of the heterogeneous account-device graph and the characteristics of accounts activities in the local structure. Moreover, to enrich the embeddings of users, Hu~\emph{et al.}~ \cite{HACUD} treat the users, merchants, devices in credit payment service as different types of nodes and their interactions as edges in a HG, and propose a meta-path-based heterogeneous graph embedding method, called HACUD, to classify the cash-out user.  Li~\emph{et al.}~ \cite{li2019spam} treat the users and items as nodes in a bipartite graph and associate the reviews as edge features to detect the spam reviews on Xianyu App. Then, a heterogeneous GNN is proposed to classify whether a review is spam or not, based on its local heterogeneous information and global context.

\begin{figure*}[htbp]
	\centering
	\subfigure[Malware detection \cite{hou2017hindroid}.]{
	\includegraphics[height=1in,width=0.22\linewidth]{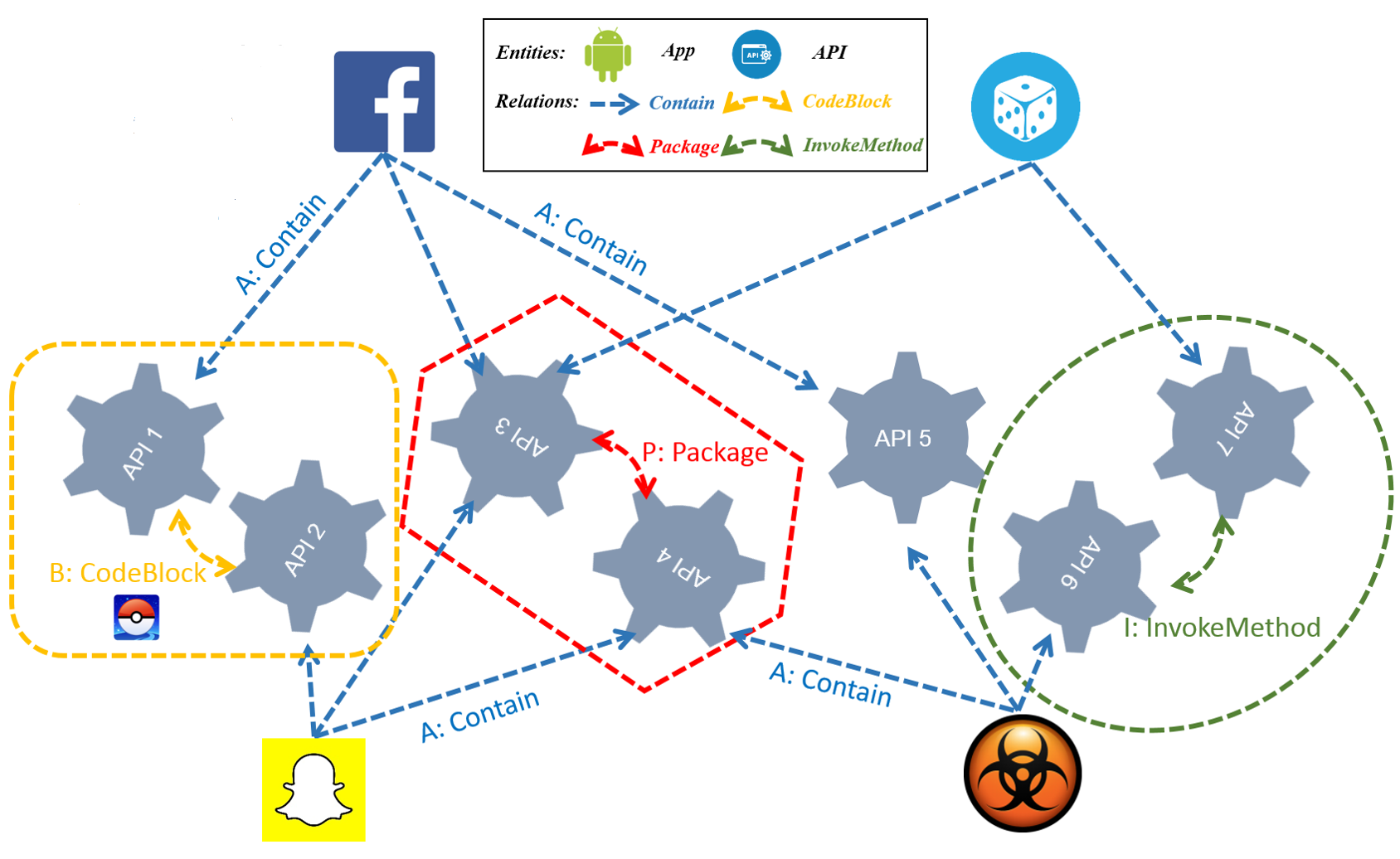}
	\label{fig:HinDroid}
	}
	\hspace{7mm}
	\subfigure[Key player identification~ \cite{zhang2019key}.]{
	\includegraphics[height=1in,width=0.28\linewidth]{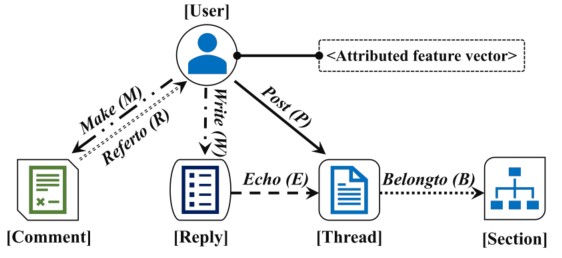}
	\label{fig:AHIN security}
		
	}
	\hspace{7mm}
	\subfigure[Drug trafficker identification \cite{zhang2019your}.]{
		\includegraphics[height=1in,width=0.35\linewidth]{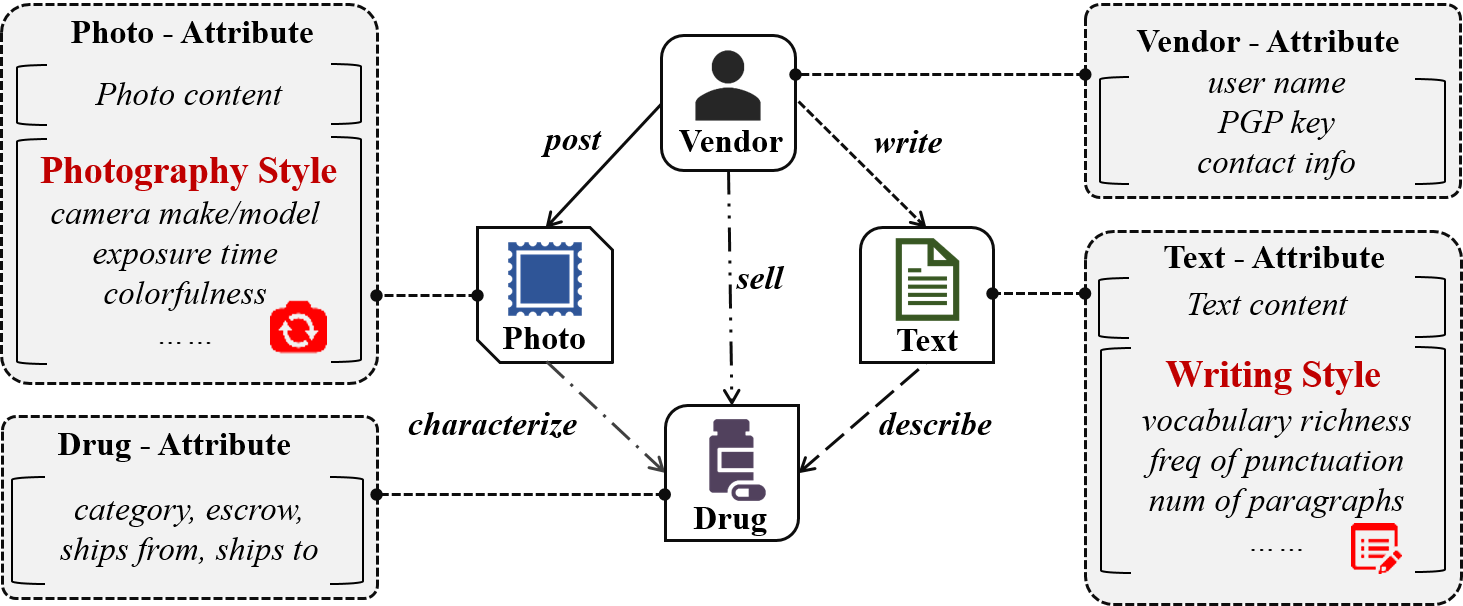}
		\label{fig:ahin}
	}
	\caption{The representative HGs in cybersecurity applications.}
	\label{fig:security}
\end{figure*}

\subsection{Cybersecurity}

Security has been one of the biggest threaten for social development, and it causes countless loss of property and lives. As multiple heterogeneous entities and complex structure are usually involved in security system, recently researchers pay more attention to use HG embedding methods to detect outliers in a wide range of security areas, such as malware detection, key player identification in underground forum, drug trafficker identification.

\par With the broad scale proliferation of increasingly interconnected devices, malware (e.g., trojans, ransomware, scamware) that deliberately fulfills the harmful intent to device users has become a major threat to compromise the security in cyberspace \cite{ye2017survey}. In particular, the explosive growth and increasing sophistication of Android malware call for new defensive techniques that are capable of protecting mobile users against novel threats \cite{felt2011survey}.
To combat the evolving Android malware attacks, HG-based methods have been proposed and applied in anti-malware industry. As shown in Fig.~\ref{fig:HinDroid}, HinDroid \cite{hou2017hindroid} was first proposed to construct a HG to model the complex relations among application programming interface (APIs) and Android applications (apps), based on which meta-paths are used to formulate the relatedness among apps and multi-kernel learning algorithm is proposed to build the classification model for malware detection. Besides modeling apps and APIs, Fan~\emph{et al.}~ \cite{fan2018gotcha} model more types of entities involved in malware into a HG, such as, file, archive and machine, and a metagraph based embedding method is designed to encode high-level semantic similarities between files. After these methods, a series of HG embedding methods are proposed for dynamic malware detection~ \cite{ye2019out}, adversarial attack and defense in malware~ \cite{hou2019alphacyber}, unknown malware detection~ \cite{wang2019unknown} and cyber threat intelligence~ \cite{gao2020hincti}.

\par Besides android malware detection, HG embedding methods also play an important role in detecting targeted objects in other security areas which have multiple types of entities and relations available. Zhang~\emph{et al.} \cite{zhang2019key} extract multiple relations from the underground forum data and construct an attributed HG (AHG) for key player identification, shown in Fig.~\ref{fig:AHIN security}. By treating the relatedness over users depicted by each meta-path as one view, a multi-view GCN is proposed to identify the key player. As illustrated in Fig.~\ref{fig:ahin}, Zhang~\emph{et al.} \cite{zhang2019your} leverage AHG to depict vendors, drugs, texts, photos and their associated attributes in darknet markets for drug trafficker identification. Then an attribute-aware AHG embedding method, named \textit{Vendor2Vec}, consisting of attribute-aware meta-path random walk and skip-gram technique, is proposed to predict whether a given pair of vendors are the same individual or not. 

\begin{table*}[htbp]
  \centering
  \caption{A summary of commonly used HG datasets.}
  \resizebox{\linewidth}{!}{
    \begin{tabular}{|c|c|c|c|c|c|c|c|c|c|c|}
    \hline
    \multirow{2}{*}{Dataset} & \multicolumn{3}{c|}{Statistics} & \multicolumn{2}{c|}{Side information} & \multicolumn{4}{c|}{Task}     & \multirow{2}{*}{Related Papers} \\
	\cline{2-10}          & Node & Link & Meta-path & Timestamp & Attribute & Node classification & Multi classification & Recommendation & Link prediction &  \\
    \hline
    \hline
    DBLP	& \begin{tabular}[c]{@{}c@{}} Author(A)\\Paper(P)\\Term(T)\\Venue(V) \end{tabular} & \begin{tabular}[c]{@{}c@{}} A-P\\P-P\\P-T\\P-V \end{tabular} & \begin{tabular}[c]{@{}c@{}} \\APA\\APAPA\\APCPA\\APTPA\\APVPA\\ \end{tabular} & $\surd$ & $\surd$ & $\surd$ & $\surd$ & & $\surd$ & \begin{tabular}[c]{@{}c@{}} \\ \cite{shi2018aspem, shi2018easing, he2019hetespaceywalk, hussein2018meta, lee2019bhin2vec} \\ \cite{gui2016large, DyHNE, fu2017hin2vec, wang2019heterogeneous, fu2020magnn} \\ \end{tabular} \\
    	

    \hline
    Aminer	& \begin{tabular}[c]{@{}c@{}} Paper(P)\\Author(A)\\Keyword(W)\\Venue(V)\\Conference(C)\\Term(T) \end{tabular} & \begin{tabular}[c]{@{}c@{}} P-A\\P-P\\P-V\\P-W\\P-T\\P-C \end{tabular} & \begin{tabular}[c]{@{}c@{}} \\APA\\WPW\\APVPA\\APTPA\\APCPA\\APWPA\\ \\ \end{tabular} & $\surd$ & $\surd$ & $\surd$ & $\surd$ & & $\surd$ & \begin{tabular}[c]{@{}c@{}} \\ \cite{MV-ACM,dong2017metapath2vec,DyHNE,zhang2019shne,zhang2019heterogeneous} \\ \cite{chen2017task,zhang2018camel,park2019task} \end{tabular} \\
    

    \hline
    Yelp	& \begin{tabular}[c]{@{}c@{}} User(U)\\Business(B)\\Compliment(Co)\\City(Ci)\\Category(Ca) \end{tabular} & \begin{tabular}[c]{@{}c@{}} U-U\\U-B\\U-Co\\B-Ci\\B-Ca\\ \end{tabular} & \begin{tabular}[c]{@{}c@{}} \\UBU\\UCoU\\UBCiBU\\UBCaBU\\BUB\\BCiB\\BCaB\\BUCoUB\\ \\\end{tabular} & $\surd$ &       & $\surd$ & $\surd$ & $\surd$ & $\surd$ & \begin{tabular}[c]{@{}c@{}} \cite{chen2018pme,he2019hetespaceywalk,gui2016large,DyHNE,shi2018heterogeneous} \\ \cite{zhao2019motif,HeteRec,fu2017hin2vec} \end{tabular} \\

    \hline
    Amazon	& \begin{tabular}[c]{@{}c@{}} User(U)\\Business(Bu)\\Category(C)\\Brand(Br)\\Aspect(A) \end{tabular} & \begin{tabular}[c]{@{}c@{}} \\U-Bu\\Bu-C\\Bu-Br\\R-Bu\\R-A\\ \\ \end{tabular} & N/A &		&  $\surd$  &       &       & $\surd$ & $\surd$ & \begin{tabular}[c]{@{}c@{}} \cite{MV-ACM,cen2019representation,zhao2019motif,zhao2019intentgc} \end{tabular} \\

    \hline
    IMDB	& \begin{tabular}[c]{@{}c@{}} User(U)\\Movie(M)\\Actor(A)\\Director(D)\\Genre(G) \end{tabular} & \begin{tabular}[c]{@{}c@{}} A-M\\U-M\\G-M\\D-M \end{tabular} & \begin{tabular}[c]{@{}c@{}} \\MUM\\MAM\\MDM\\MGM\\UMU\\UMAMU\\UMDMU\\UMGMU\\ \\ \end{tabular} & & $\surd$ & $\surd$ & & $\surd$ & $\surd$ & \begin{tabular}[c]{@{}c@{}} \cite{shi2018aspem,wang2019heterogeneous,fu2020magnn} \end{tabular} \\

    \hline
    Douban	& \begin{tabular}[c]{@{}c@{}} User(U)\\Movie(M)\\Group(G)\\Location(L)\\Direction(D)\\Actor(A)\\Type(T) \end{tabular} & \begin{tabular}[c]{@{}c@{}} U-U\\U-G\\U-M\\U-L\\M-D\\M-T\\M-A \end{tabular} & \begin{tabular}[c]{@{}c@{}} \\MUM\\MTM\\MDM\\MAM\\UMU\\UMAMU\\UMDMU\\UMTMU\\ \\ \end{tabular} & & & $\surd$ & $\surd$ & $\surd$ & $\surd$ & \begin{tabular}[c]{@{}c@{}} \cite{he2019hetespaceywalk,lee2019bhin2vec,shi2018heterogeneous} \end{tabular} \\

    \hline
    \end{tabular}}
  \label{datasets}
\end{table*}

\subsection{Others}


With the development of biological medicine, medical informatics has received considerable attentions, especially, mining Electronic Health Records (EHR) for reducing error and improving quality of disease diagnosis \cite{medical2020cao}. Previous work on medical HG mainly utilizes HeteSim~ \cite{shi2014hetesim} to analyze the similarities between objects~ \cite{xiao2017prediction}. Recently, Hosseini~\emph{et al.}~ \cite{anahita2018heteromed} treat the diagnostic and treatment events as the nodes and corresponding relations extracted from raw text as edges in a HG, and propose a meta-path-guided HG embedding method to rank each patient's potential diagnosis.

\par Besides, heterogeneous graph embedding is also applied in real-time event prediction on ride-hailing platform, such as Uber\footnote{www.uber.com} and
DiDi\footnote{www.didiglobal.com}. Luo~\emph{et al.}~ \cite{luo2020dynamic} dynamically construct heterogeneous graph for each ongoing event, such as PreView page and request, to encode the attributes of the event and the condition information from its surrounding area. And a multilayer GNN is proposed to learn the impact of historical
actions and the surrounding environment on the current events, and generate an event embedding to improve the accuracy of the response model. Hong~\emph{et al.} ~ \cite{hong2020heteta} propose HetETA to leverage HG to model the spatiotemporal information in time-of-arrival (ETA) estimation task. And a multi-component GNN is proposed to model temporal information from different time spans for ETA task.

\section{Benchmarks and Open-source Tools}
In this section, we summarize the commonly used datasets of heterogeneous graph embedding. Besides, we introduce some useful resources and open-source tools about heterogeneous graph embedding.

\subsection{Benchmark Datasets}
High-quality datasets are essential for academic research. Here, we introduce some popular real world HG datasets, which can be divided into three categories: academic networks, business networks and film networks. Specifically, we summarize their detailed statistical information in Table \ref{datasets}, including node types, link types and meta-paths etc.

\begin{itemize}
	\item \textbf{DBLP}\footnotemark[2] This is a network that reflects the relationship between authors and papers. There are four types of nodes: author, paper, term and venue.
	\item \textbf{Aminer}\footnotemark[3] This academic network is similar to DBLP, but with two additional node types: keyword and conference.
	\item \textbf{Yelp}\footnotemark[4] This is a social media network, including five types of nodes: user, business, compliment, city and category.
	\item \textbf{Amazon}\footnotemark[5] This is an E-commercial network, which records the interactive information between users and products, including co-viewing, co-purchasing, etc.
	\item \textbf{IMDB}\footnotemark[6] This is a film rating network, recording the preferences of users on different films. Each film contains its directors, actors and genre.
	\item \textbf{Douban}\footnotemark[7] This network is similar to IMDB, but it contains more user information, such as the group and location of the users.
\end{itemize}






\footnotetext[2]{http://dblp.uni-trier.de}
\footnotetext[3]{https://www.aminer.cn}
\footnotetext[4]{http://www.yelp.com/dataset challenge/}
\footnotetext[5]{http://jmcauley.ucsd.edu/data/amazon}
\footnotetext[6]{https://grouplens.org/datasets/movielens/100k/}
\footnotetext[7]{http://movie.douban.com/}

\subsection{Open-source Code and Tools}
Open resources and tools are of great significance to the development of academic research. In this subsection, we provide the resources of heterogeneous graph embedding and introduce some useful open-source platforms and toolkits.

\begin{table*}[t]
  \centering
  \caption{Source code of related papers.}
    \begin{tabular}{l|l|c}
    \hline
    Method & Source code & Programing platform\\
    \hline
    AspEM \cite{shi2018aspem} & https://github.com/ysyushi/aspem & Python\\
    HEER \cite{shi2018easing}  & https://github.com/GentleZhu/HEER & Python\\
    BHIN2vec \cite{lee2019bhin2vec} & https://github.com/sh0416/BHIN2VEC & Pytorch \\
    HEBE \cite{gui2016large}  & https://github.com/olittle/Hebe & C++ \\
    DyHNE \cite{DyHNE} & https://github.com/rootlu/DyHNE & Python \& Matlab \\
    HIN2vec \cite{fu2017hin2vec} & https://github.com/csiesheep/hin2vec & Python \& C++\\
    HAN \cite{wang2019heterogeneous}   & https://github.com/Jhy1993/HAN & Tensorflow \\
    MAGNN \cite{fu2020magnn} & https://github.com/cynricfu/MAGNN & Pytorch \\
    
	metapath2vec \cite{dong2017metapath2vec} & https://github.com/apple2373/metapath2vec & Tensorflow \\
    SHNE \cite{zhang2019shne}  & https://github.com/chuxuzhang/WSDM2019\_SHNE & Pytorch \\
    HetGNN \cite{zhang2019heterogeneous} & https://github.com/chuxuzhang/KDD2019\_HetGNN & Pytorch\\
    TaPEm \cite{park2019task} & https://github.com/pcy1302/TapEM & Python\\
	HeRec \cite{shi2018heterogeneous} & https://github.com/librahu/HERec & Python \\
	FMG \cite{zhao2019motif}   & https://github.com/HKUST-KnowComp/FMG & Python \& C++ \\
	HeteRec \cite{HeteRec}	& https://github.com/mukulg17/HeteRec & R\\
    GATNE \cite{cen2019representation} & https://github.com/THUDM/GATNE & Pytorch \\
    IntentGC \cite{zhao2019intentgc}  & https://github.com/peter14121/intentgc-models & Python \\
    \hline
    \end{tabular}
  \label{codes}
\end{table*}

\subsubsection{Open-source Code}
Source code is important for researchers to reproduce the corresponding method. In Table \ref{datasets}, we refer to the related papers of the datasets. Furthermore, we collect the source code of the related papers and list them in Table \ref{codes}. Besides, we provide some commonly used website about graph embedding.
\begin{itemize}
    \item Stanford Network Analysis Project (SNAP). It is a network analysis and graph mining library, which contains different types of networks and multiple network analysis tools. The address is http://snap.stanford.edu/.
    \item ArnetMiner (AMiner) \cite{AMiner}. In the early days, it was an academic network used for data mining. Now it becomes to a comprehensive academic system that provides a variety of academic resources. The address is https://www.aminer.cn/.
    \item Open Academic Society (OAS). It is an open and expanding knowledge graph for research and education, contributed by Microsoft Research and AMiner. It publishes Open Academic Graph (OAG), which unifies two billion-scale academic graphs. The address is https://www.openacademic.ai/.
    \item HG Resources. It is a website focusing on heterogeneous graphs, which collects a series of papers on HG and divides them into different categories, including classficiation, clustering and embedding. Code and datasets of the popular methods are also provided. The address is http://shichuan.org/.
\end{itemize}

\subsubsection{Available Tools}
Open-source platforms and toolkits can help researchers build the workflow of graph embedding quickly and easily. Generally, there are many toolkits designed for homogeneous graph. For example, OpenNE\footnotemark[8] and CogDL\footnotemark[9]. However, the toolkits and platforms for heterogeneous graph are rarely mentioned. To bring this gap, we summary the popular toolkits and platforms that are suitable for heterogeneous graph.

\footnotetext[8]{https://github.com/thunlp/OpenNE}
\footnotetext[9]{https://github.com/THUDM/cogdl}

\begin{itemize}
    \item AliGraph. It is an industrial-grade machine learning platform for graph data, supporting the calculation of hundreds of millions of nodes and edges. Besides, it considers the characteristics of real world industrial graph data, i.e., large-scale, heterogeneous, attributed and dynamic, and makes special optimizations. One instance can be found in https://www.aliyun.com/product/bigdata/product.
    \item Deep Graph Library (DGL). It is an open-source deep learning platform for graph data, which designs its own data structures and implements many popular methods. Specifically, it provides independent Application Programming Interfaces (APIs) for homogeneous graph, heterogeneous graph and knowledge graph. One instance can be found in https://www.dgl.ai/.
    \item Pytorch Geometric. It is a geometric deep learning extension library for pytorch. Specifically, it focuses on the methods for deep learning on graphs and other irregular structures. Same as DGL, it also has its own data structures and operators. One instance can be found in https://pytorch-geometric.readthedocs.io/en/latest/.
    \item OpenHINE. It is an open-source toolkit for heterogeneous graph embedding, which implements many popular heterogeneous graph embedding methods with a unified data interface. One instance can be found in  https://github.com/BUPT-GAMMA/OpenHINE.
\end{itemize}

\section{Challenges and Future Directions}
Heterogeneous graph embedding has made great progress in recent years, which clearly shows that it is a powerful and promising graph analysis paradigm. In this section, we discuss additional issues/challenges and explore a series of possible future research directions.

\subsection{Preserving HG Structures}
The basic success of heterogeneous graph embedding builds on the HG structure preservation. This also motivates many heterogeneous graph embedding methods to exploit different HG structures, where the most typical one is meta-path \cite{dong2017metapath2vec, shi2016survey}. Following this line, meta-graph structure is naturally considered \cite{zhang2018metagraph2vec}. However, HG is far more than these structures. Selecting the most appropriate meta-path is still very challenging in the real world. An improper meta-path will fundamentally hinder the performance of heterogeneous graph embedding method. Whether we can explore other techniques, e.g., motif \cite{zhao2019motif, huang2016meta} or network schema \cite{NSHE} to capture HG structure is worth pursuing. Moreover, if we rethink the goal of traditional graph embedding, i.e., replacing the structure information with the distance/similarity in a metric space, a research direction to explore is whether we can design a heterogeneous graph embedding method which can naturally learn such distance/similarity rather than using pre-defined meta-path/meta-graph.

\subsection{Capturing HG Properties}
As mentioned before, many current heterogeneous graph embedding methods mainly take the structures into account. However, some properties, which usually provide additional useful information to model HG, have not been fully considered. One typical property is the dynamics of HG, i.e., a real world HG always evolves over time. Despite that the incremental learning on dynamic HG is proposed \cite{DyHNE}, dynamic heterogeneous graph embedding is still facing big challenges. For example, \cite{change2vec} is only proposed with a shallow model, which greatly limits its embedding ability. How can we learn dynamic heterogeneous graph embedding in deep learning framework is worth pursuing. The other property is the uncertainty of HG, i.e., the generation of HG is usually multi-faceted and the node in a HG contains different semantics. Traditionally, learning a vector embedding usually cannot well capture such uncertainty. Gaussian distribution may innately represent the uncertainty property \cite{VGAE, zhu2018variational}, which is largely ignored by current heterogeneous graph embedding methods. This suggests a huge potential direction for improving heterogeneous graph embedding.

\subsection{Deep Graph Learning on HG Data}
We have witnessed the great success and large impact of GNNs, where most of the existing GNNs are proposed for homogeneous graph \cite{kipf2017semi, velickovic2018graph}. Recently, HGNNs have attracted considerable attention \cite{wang2019heterogeneous, zhang2019heterogeneous, fu2020magnn, cen2019representation}.

One natural question may arise that what is the essential difference between GNNs and HGNNs. More theoretical analysis on HGNNs are seriously lacking. For example, it is well accepted that the GNNs suffer from over-smoothing problem \cite{deeper}, so will heterogeneous GNNs also have such problem? If the answer is yes, what factor causes the over-smoothing problem in HGNNs since they usually contain multiple aggregation strategies \cite{wang2019heterogeneous, zhang2019heterogeneous}.

In addition to theoretical analysis, new technique design is also important. One of the most important directions is the self-supervised learning. It uses the pretext tasks to train the neural networks, thus reducing the dependence on manual labels. \cite{self-supervised}. 
Considering the actual demand that label is insufficient, self-supervised learning can greatly benefit the unsupervised and semi-supervised learning, and has shown remarkable performance on homogeneous graph embedding \cite{DGI, M3S, peng2020self, you2020self}. Therefore, exploring self-supervised learning on heterogeneous graph embedding is expected to further facilitate the development of this area.

Another important direction is the pre-training of HGNNs \cite{GPT-GNN, GCC}. Nowadays, HGNNs are designed independently, i.e., the proposed method usually works well for some certain tasks, but the transfer ability across different tasks is ill-considered. When dealing with a new HG or task, we have to train a heterogeneous graph embedding method from scratch, which is time-consuming and requires large amounts of labels. In this situation, if there is a well pre-trained HGNN with strong generalization that can be fine-tuned with few labels, the time and label consumption can be reduced.

\subsection{Making HG embedding reliable}

Except from the properties and techniques in HG, we are also concerned about the ethical issues in HG embedding, such as fairness, robustness and interpretability. Considering that most methods are black boxes, making HG embedding reliable is an important future work.

\textbf{Fair HG embedding.} The embeddings learned by methods are sometimes highly related to certain attributes, e.g., age or gender, which may amplify the societal stereotypes in the prediction results \cite{fair2019bose, fairsurvey2019du}. Therefore, learning fair or de-biased embeddings is an important research direction. There are some researches on the fairness of homogeneous graph embedding \cite{fair2019bose, fairwalk2019rahman}. However, the fairness of HG is still an unsolved problem, which is an important research direction in the further.

\textbf{Robust HG embedding.} Also, the robustness of  HG embedding, especially the adversarial attacking, is always an important problem \cite{attack}. Since many real world applications are built based on HG, the robustness of HG embedding becomes an urgent yet unsolved problem. What is the weakness of HG embedding and how to enhance it to improve the robustness need to be further studied.

\textbf{Explainable HG embedding.} Moreover, in some risk aware scenarios, e.g., fraud detection \cite{HACUD} and bio-medicine \cite{medical2020cao} , the explanation of models or embeddings is important. A significant advantage of HG is that it contains rich semantics, which may provide eminent insight to promote the explanation of heterogeneous GNNs. Besides, the emerging disentangled learning \cite{disentangle2017nara, disentangle2019ma}, which divides the embedding into different latent spaces to improve the interpretability, can also be considered.


\subsection{Technique Deployment in Real-world Applications}

Many HG-based applications have stepped into the era of graph embedding. This survey has demonstrated the strong performance of heterogeneous graph embedding methods on E-commerce and cybersecurity. Exploring more capacity of heterogeneous graph embedding on other areas holds great potential in the future. For example, in software engineering area, there are complex relations among test sample, requisition form, and problem form, which can be naturally modeled as HG. Therefore, heterogeneous graph embedding is expected to open up broad prospects for these new areas and become promising analytical tool. Another area is the biological systems, which can also be naturally modeled as a HG. A typical biological system contains many types of objects, e.g., Gene Expression, Chemical, Phenotype, and Microbe. There are also multiple relations between Gene Expression and Phenotype \cite{koki2017bio}. HG structure has been applied to biological system as an analytical tool, implying that heterogeneous graph embedding is expected to provide more promising results. 

In addition, since the complexity of HGNNs are relatively large and the techniques are difficult to parallelize, it is difficult to apply the existing HGNNs to large-scale industrial scenarios. For example, the number of nodes in E-commerce recommendation may reach one billion \cite{zhao2019intentgc}. Therefore, successful technique deployment in various applications while resolving the scalability and efficiency challenges will be very promising.

\subsection{Others}

Last but not least, there are also some important future work that cannot be summarized in the previous sections. Therefore, we carefully discuss them in this subsection.

\textbf{Hyperbolic heterogeneous graph embedding.} Some recent researches point out that the underlying latent space of graph may be non-Euclidean, but in hyperbolic space \cite{hyperbolic}. Some attempts have been made towards hyperbolic graph/heterogeneous graph embedding, and the results are rather promising \cite{hyperGNN1, hyperGNN2, wang2019hyperbolic}. However, how to design an effective hyperbolic heterogeneous GNNs is still challenging, which can be another research direction. 

\textbf{Heterogeneous graph structure learning.} Under the current heterogeneous graph embedding framework, HG is usually constructed beforehand, which is independent on the heterogeneous graph embedding. This may result in that the input HG is not suitable for the final task. HG structure learning can be further integrated with heterogeneous graph embedding, so that they can promote each other. 

\textbf{Connections with knowledge graph.} Knowledge graph embedding has great potential on knowledge reasoning \cite{pan2020knowledge}. However, knowledge graph embedding and heterogeneous graph embedding are usually investigated separately. Recently, knowledge graph embedding has been successfully applied to other areas, e.g., recommender system \cite{guo2020knowledge, wang2019knowledge}. It is worth studying that how to combine knowledge graph embedding with heterogeneous graph embedding, and incorporate knowledge into heterogeneous graph embedding.

\section{Conclusion}

Heterogeneous graph embedding has significantly facilitated the HG analysis and related applications. This survey
conducts a comprehensive study of the state-of-the-art heterogeneous graph embedding methods. Thorough discussions and summarization of the
reviewed methods, along with the widely used benchmarks and resources, are systematically presented. We hope that  this survey can
provide a clean sketch on heterogeneous graph embedding, which could help both the interested readers as well as the
researchers that wish to continue working in this area.

\section{Acknowledgment}

C. Shi, X. Wang, D. Bo and S. Fan’s work is partially supported by the National Natural Science Foundation of China (No. U20B2045, 61702296, 61772082, 62002029), Meituan-Dianping Group and BUPT Excellent Ph.D. Students Foundation (No. CX2020115, CX2019127).
Y. Ye's work is partially supported by the NSF under grants IIS-1951504, CNS-2034470, CNS-1940859, CNS-1814825 and OAC-1940855, the DoJ/NIJ under grant NIJ 2018-75-CX-0032.
P. S. Yu's work is supported in part by NSF under grants III-1763325, III-1909323, and SaTC-1930941.



%

%



\ifCLASSOPTIONcaptionsoff
  \newpage
\fi



%

%
%

\bibliographystyle{IEEEtran}
\bibliography{HINRL.bib}

%








\end{document}